\documentclass[fleqn,usenatbib]{mnras}

\usepackage{newtxtext,newtxmath}
\usepackage[T1]{fontenc}

\DeclareRobustCommand{\VAN}[3]{#2}
\let\VANthebibliography\thebibliography
\def\thebibliography{\DeclareRobustCommand{\VAN}[3]{##3}\VANthebibliography}

\usepackage{graphicx}	% Including figure files
\usepackage{amsmath}	% Advanced maths commands
\usepackage{ulem}
\usepackage[dvipsnames]{xcolor}

\title[Semi-Supervising Lensed Quasars]{Semi-Supervised Learning for Lensed Quasar Detection}

\author[D. Sweeney et al.]{
David Sweeney,$^{1, 2}$\thanks{E-mail: davidsweeneyastro@gmail.com}
Alberto Krone-Martins,$^{2}$
Daniel Stern$^{3}$,
Peter Tuthill$^{1}$,
Richard Scalzo$^{4}$,
\newauthor
\hspace{1mm}George Djorgovski$^{5}$,
Christine Ducourant$^{6}$,
Ashish Mahabal$^{5}$,
Ramachrisna Teixeira$^{7}$
\newauthor
\hspace{1mm}and Matthew Graham$^{5}$
\\
$^{1}$Sydney Institute for Astronomy (SIfA), The University of Sydney, Physics Road, Sydney 2050, Australia\\
$^{2}$Donald Bren School of Information and Computer Sciences, University of California, Irvine, CA 92697, USA\\
$^{3}$Jet Propulsion Laboratory, California Institute of Technology, 4800 Oak Grove Drive, Pasadena, CA 91109, USA\\
$^{4}$CSIRO Data61, Clayton, VIC 3168, Australia\\
$^{5}$California Institute of Technology, 1200 E. California Blvd, Pasadena, CA 91125, USA\\
$^{6}$Laboratoire d’Astrophysique de Bordeaux, Univ. Bordeaux, CNRS, B18N, allée Geoffroy Saint-Hilaire, 33615 Pessac, France\\
$^{7}$Instituto de Astronomia, Geofísica e Ciências Atmosféricas, Universidade de São Paulo, Rua do Matão, 1226, Cidade Universitária, 05508-900 São Paulo, SP, Brazil
}

\date{Accepted XXX. Received YYY; in original form ZZZ}

\pubyear{2015}

\begin{document}
\label{firstpage}
\pagerange{\pageref{firstpage}--\pageref{lastpage}}
\maketitle

% Abstract of the paper
\begin{abstract}
Lensed quasars are key to many areas of study in astronomy, offering a unique probe into the intermediate and far universe. However, finding lensed quasars has proved difficult despite significant efforts from large collaborations. These challenges have limited catalogues of confirmed lensed quasars to the hundreds, despite theoretical predictions that they should be many times more numerous. 
We train machine learning classifiers to discover lensed quasar candidates. By using semi-supervised learning techniques we leverage the large number of potential candidates as unlabelled training data alongside the small number of known objects, greatly improving model performance.
We present our two most successful models: (1) a variational autoencoder trained on millions of quasars to reduce the dimensionality of images for input to a dense neural network classifier that can make accurate predictions and (2) a convolutional neural network trained on a mix of labelled and unlabelled data via virtual adversarial training. These models are both capable of producing high-quality candidates, as evidenced by our discovery of GRALJ140833.73+042229.98. The success of our classifier, which uses only multi-band images, is particularly exciting as it can be combined with existing classifiers, which use other data than images, to improve the classifications of both models and discover more lensed quasars.
\end{abstract}

% Select between one and six entries from the list of approved keywords.
% Don't make up new ones.

%%%%%%%%%%%%%%%%%%%%%%%%%%%%%%%%%%%%%%%%%%%%%%%%%%

%%%%%%%%%%%%%%%%% BODY OF PAPER %%%%%%%%%%%%%%%%%%

\section{Introduction}
\label{sec:difficulties}
\begin{figure}
    \centering
    \includegraphics[width=\columnwidth]{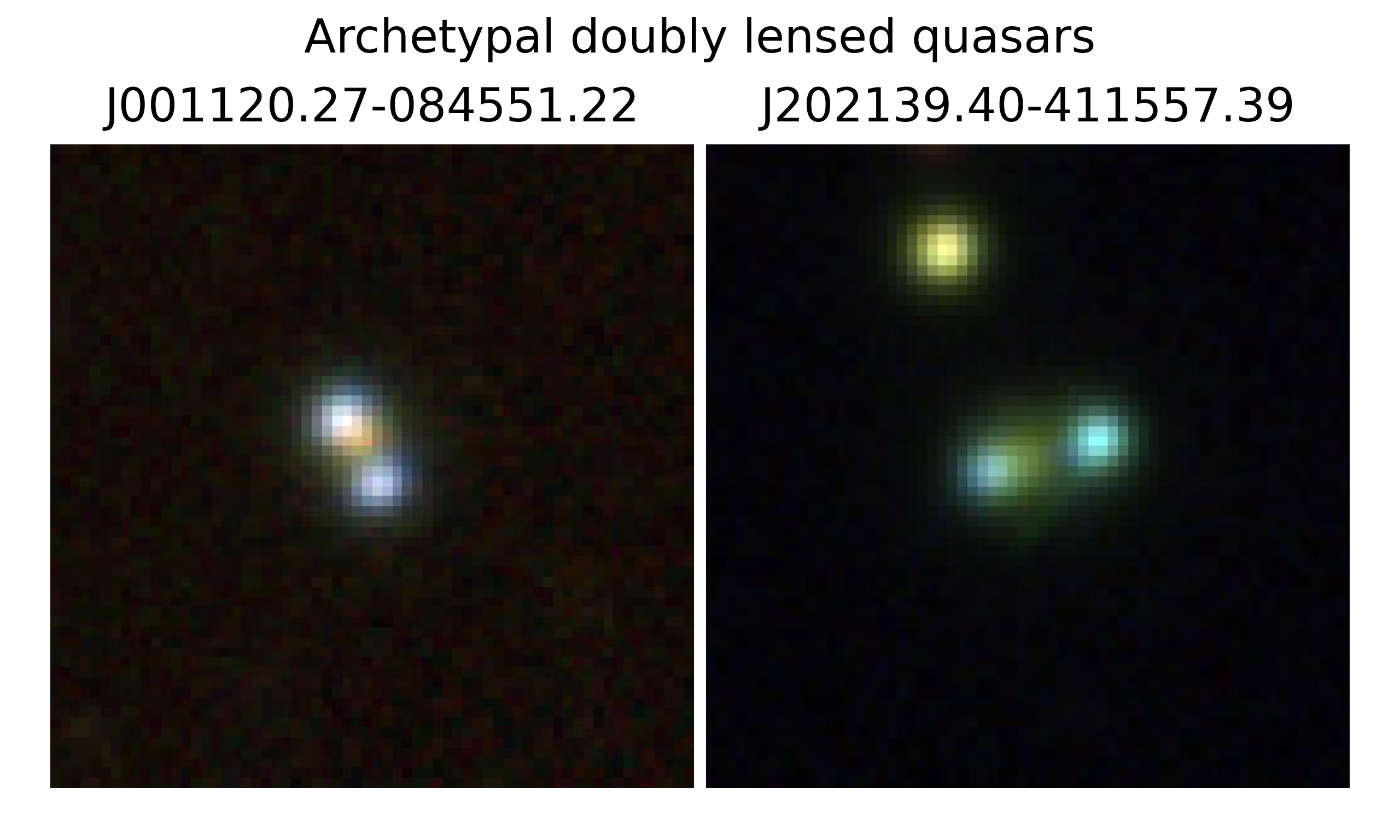}
    \includegraphics[width=\columnwidth]{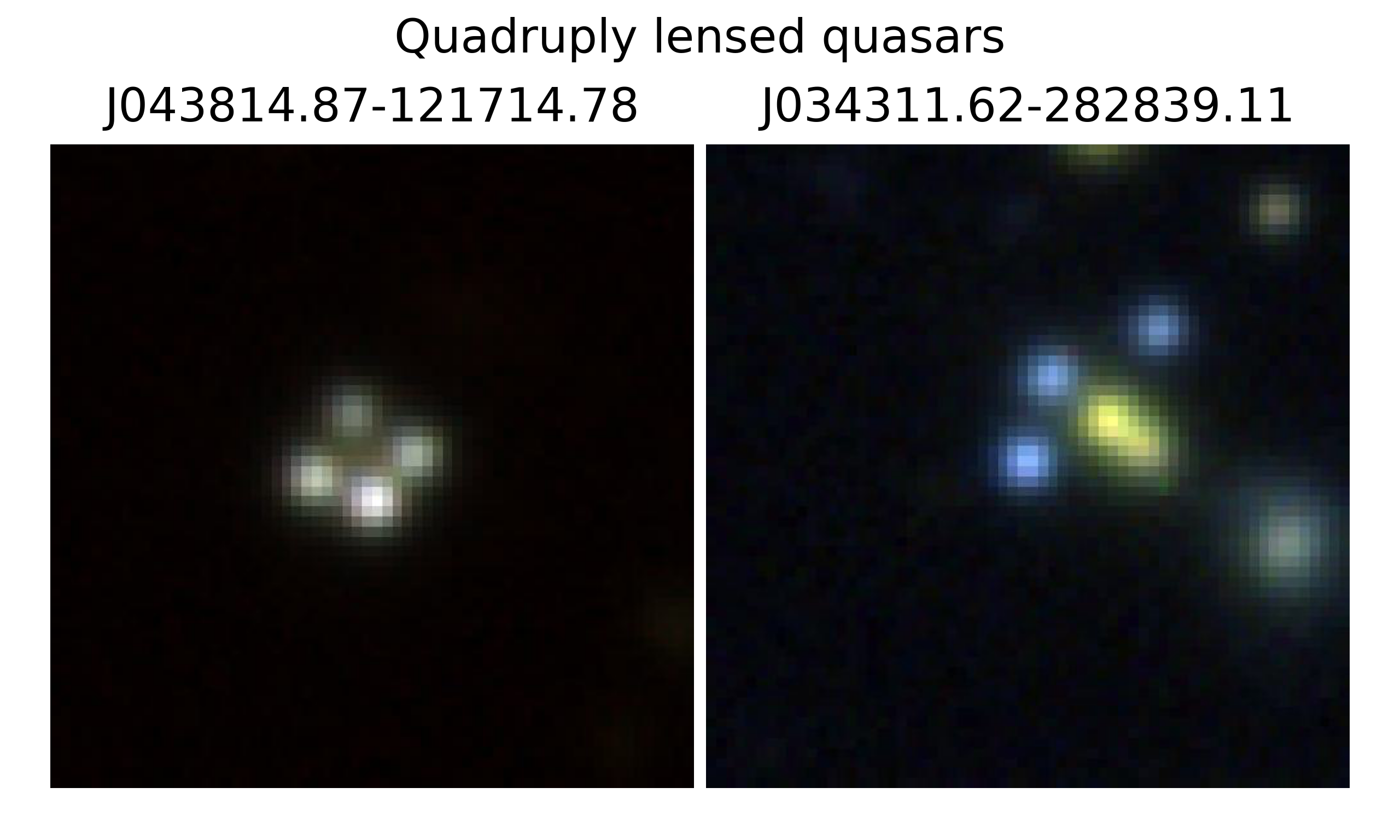}
    \includegraphics[width=\columnwidth]{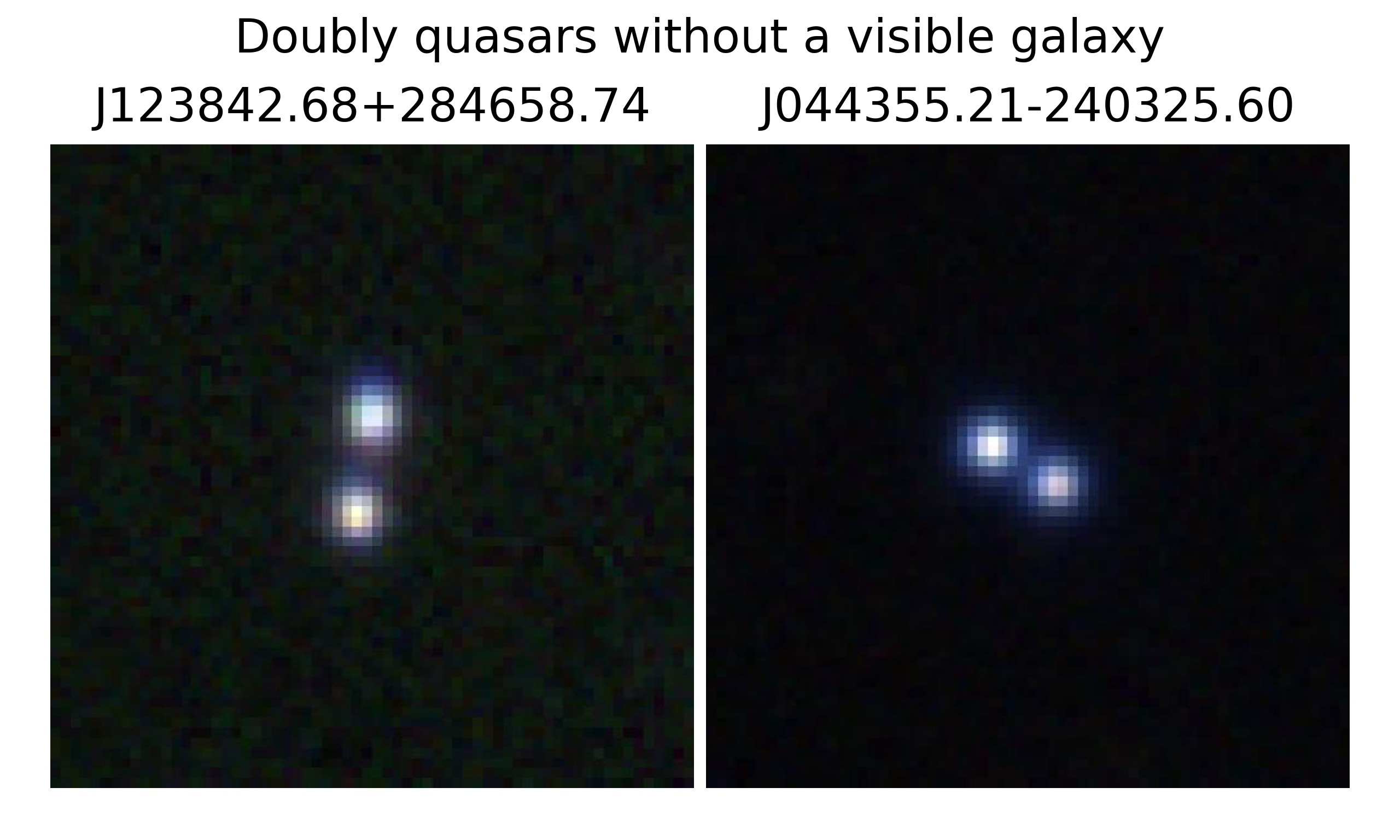}
    \caption{Six images of lensed quasars showing data from DESI. The top two images show the archetypal lensed quasar as it would be described to a student: two identically blue-white quasars on either side of a red lensing galaxy, which is often extended. This typical difference in colouring between the quasar and lensing galaxy is the reason multi-band images are so crucial --- to differentiate between other astrophysical phenomena. The middle two images show examples of quadruply lensed quasars, four blue-white quasars arranged in a kite. Sometimes, as in the left image, the lensing galaxy is barely or not at all visible and other times, as in the right image, the lensing galaxy obscures one of the quasar images. The bottom two images show more typical lensed quasars, two blue-white quasars, often with the lensing galaxy too dim to be seen and commonly one of the quasar images is reddened, presumably by the lensing galaxy. Note that the images in this paper are comprised of the $g$, $r$ and $i$ bands and visualised in images as the r, g, b channels of a standard image. Unless otherwise specified, the images are 16x16 arcsec of sky.}
    \label{fig:lens-examples}
\end{figure}

Quasars are supermassive black holes at the centres of distant galaxies that convert the energy of accreting matter efficiently to radiation, making them among the most luminous objects in the universe.
Massive galaxies along our line of sight to a distant quasar warp the intervening space-time, leading to multiple images of the same lensed quasar in a phenomena called gravitational lensing \citep{Einstein1916, Einstein1936, Zwicky1937}. The alignment of the quasar, galaxy and observer as well as the distribution of the galaxy's mass can combine to result in two, or in rare cases, four observable images of the quasar, as depicted in Figure~\ref{fig:lens-examples}. 

Since their first discovery by \citet{Walsh1979}, lensed quasars have been hotly sought after astronomical objects as probes of cosmology and the structure of galaxies and their surrounds. Because the light from each image of the lensed quasar has travelled different paths through space-time, although originating at a single source, we can compare the observed spectra of the different images to infer properties of the circumgalactic medium of the lensing galaxy \citep[e.g. ][]{Rauch2001, Rauch2011}, the intergalactic medium \citep[e.g. ][]{Foltz1984, Smette1995} and even the Hubble constant \citep[e.g. ][]{Refsdal1964, Meyer2023}. 

Beyond exploiting the differences in the lightpaths experienced by each quasar image, gravitational lensing allows us to peer more deeply into the distant universe. The gravitational lensing magnifies the distant quasar, allowing us to observe quasars which otherwise would have been too faint to be detected by our telescopes. Given the apparent link between the formation of supermassive black holes and their host galaxies, identifying the largest redshift at which quasars are active is a critical test for our theories of galaxy formation \citep{Volonteri2010, Weinberger2018}. By identifying lensed quasars we are able to boost the sensitivity of our instruments, potentially probing higher redshifts than observing limits would otherwise allow.

To reliably identify quasars several challenges must be overcome. Firstly, quasars themselves are rare, appearing in only a small fraction of galaxies; lensed quasars are many times rarer, occurring at a rate of about $1/1000$--$1/10\,000$ quasars. This results in a class imbalance for any manual or automated classification effort.

The second difficulty is the small number of confirmed lensed quasars, now sitting at about 250 \citep[e.g.][]{Ducourant2018, Lemon2019, Krone-Martins2019}, with perhaps 400 more which have been identified but as yet are unpublished. 
Compared to modern image recognition tasks, which often utilise billions of images, this number is painfully small; even the number of known quasars (which may or may not be lensed) only sits at around 6 million \citep{Milliquas8,2023Coryn}.
Labelling more data is also prohibitively expensive: to confirm whether an object is a lensed quasar we require a 3.5+~m telescope operated for hours by a highly skilled astronomer, which can still lead to an inconclusive result. Thus, while a handful of such observations can be performed each year, in the short term our set of lensed quasars will remain of size $\sim$650.

\begin{figure}
    \centering
    \includegraphics[width=\columnwidth]{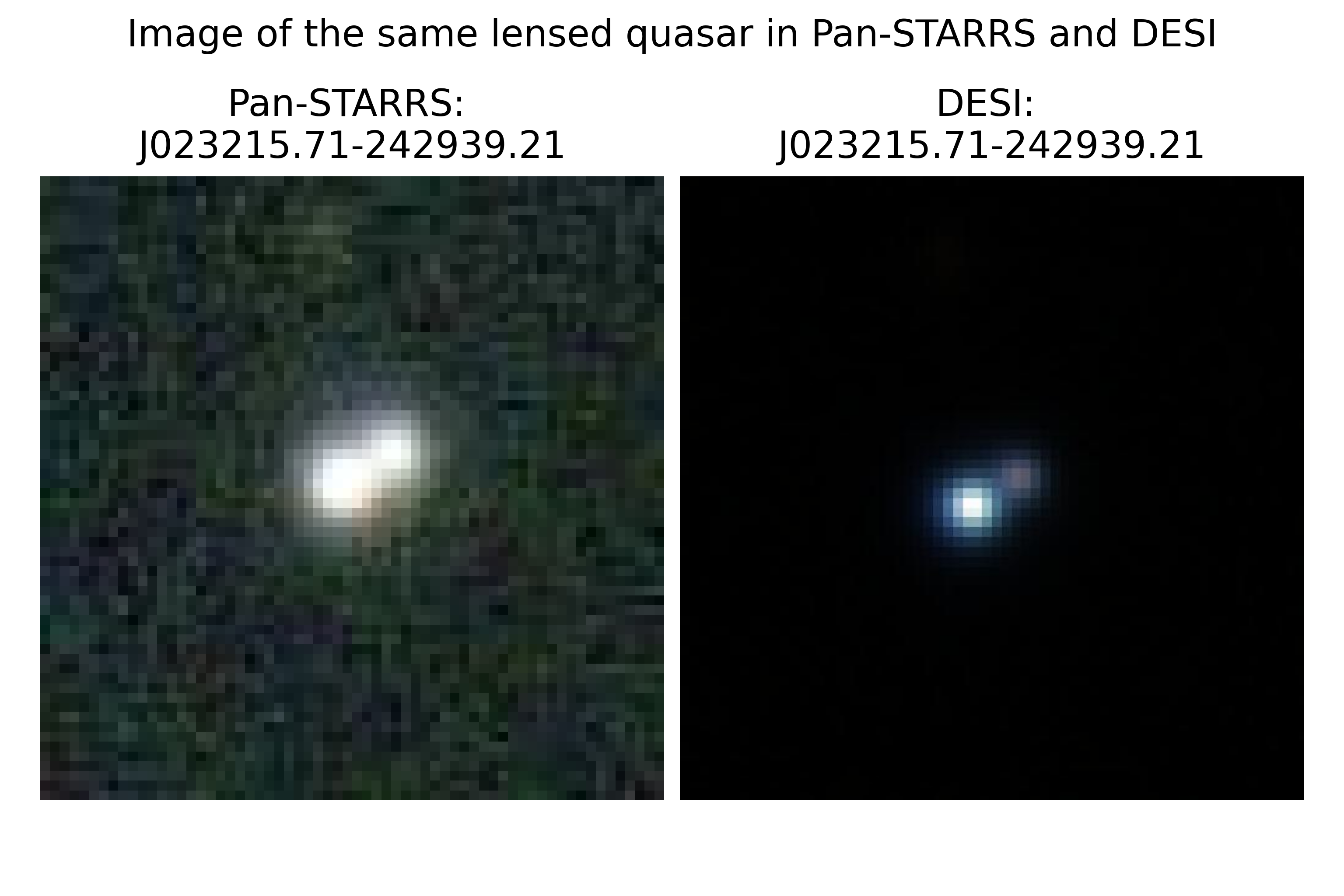}
    \\
    \vspace{-3mm}
    \includegraphics[width=\columnwidth]{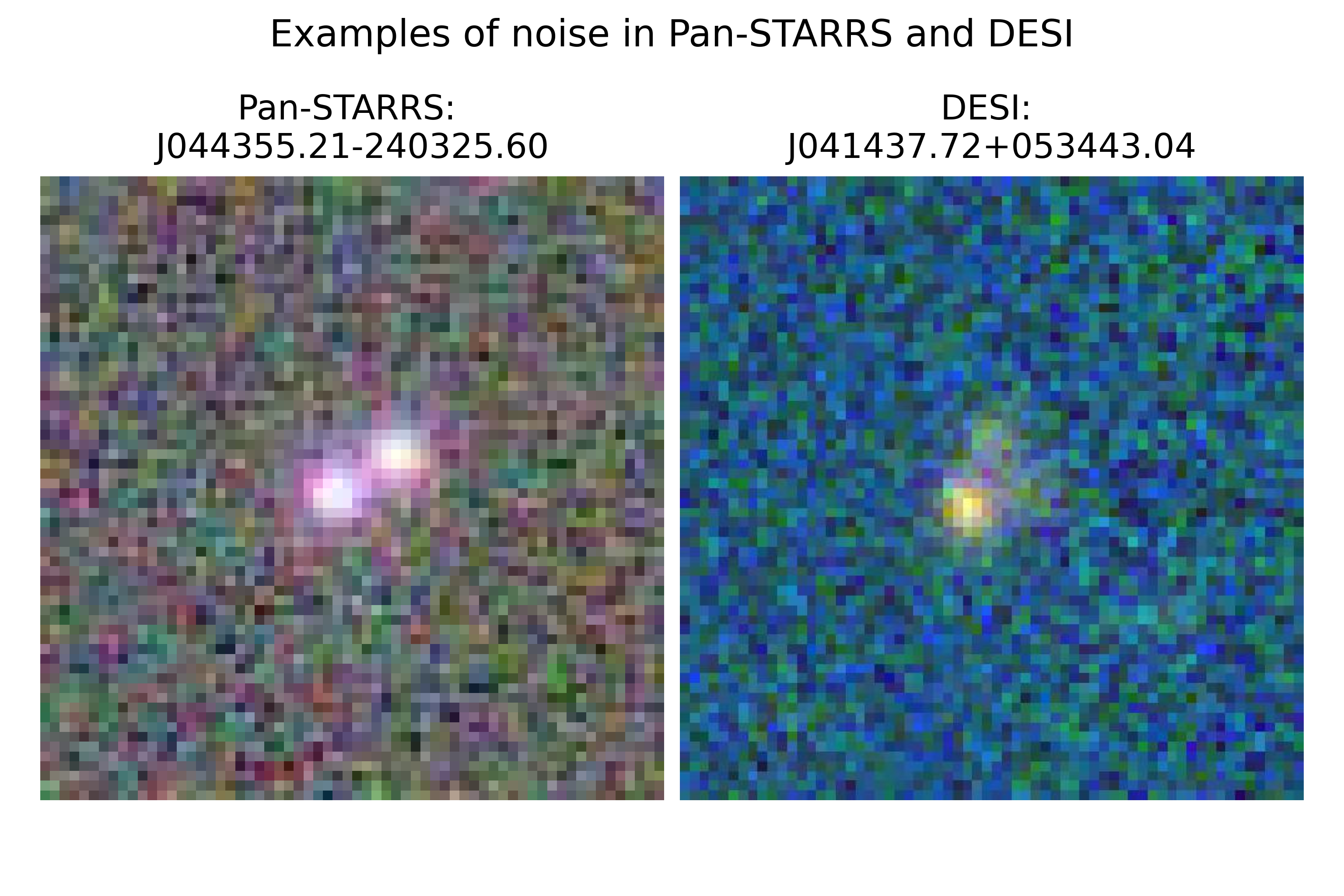}
    \\
    \vspace{-3mm}
    \includegraphics[width=\columnwidth]{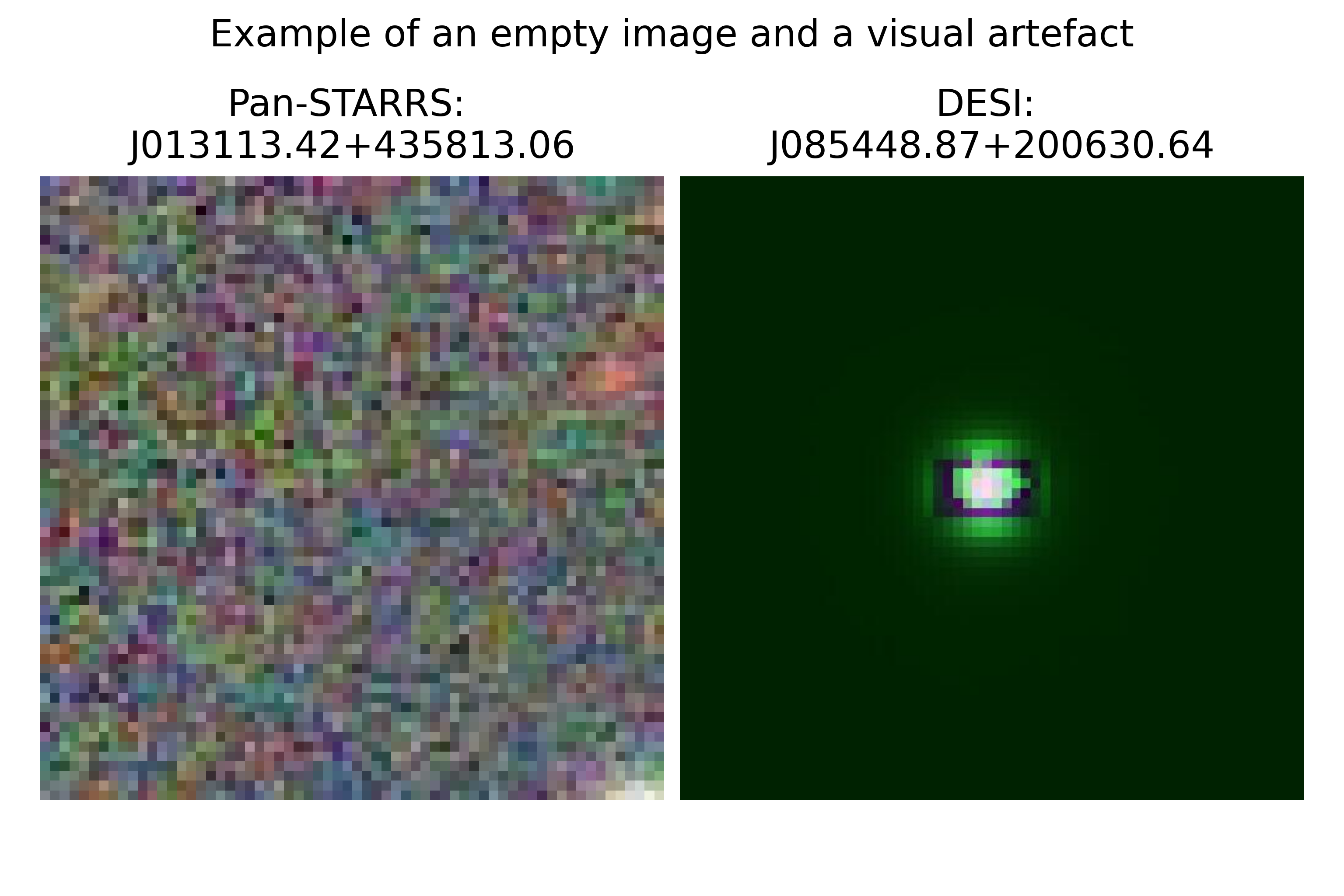}
    \caption{Six images of lensed quasars showing data from Pan-STARRS and DESI. The top two images show the same lensed quasar imaged by both Pan-STARRS and DESI surveys. The middle two images show the kind of noise evident in Pan-STARRS and DESI images. The bottom left image shows an case where no objects are visible in a Pan-STARRS image. The bottom right image shows a typical example of a visual artefact that you may see in DESI images.}
    \label{fig:image-examples}
\end{figure}

Another difficulty is the quality of the available data: not all of the objects listed in catalogues like Milliquas \citep{Milliquas8} are quasars. ``Quasars'' in catalogues like Milliquas have often been identified by applying machine learning to large surveys and, although often the best course of action, leads to mislabelled objects. Images of these quasars can be taken from the Panoramic Survey Telescope and Rapid Response System \citep[Pan-STARRS;][]{PanstarrsMain2016} survey (covering the northern sky) and/or the Dark Energy Spectroscopic Instrument \citep[DESI;][]{DESI} Legacy Imaging Surveys, covering the southern sky. Image quality and appearance differ between these two surveys and visual artefacts can further complicate classification of lensed quasars. Compared to image classification tasks in other domains, the images of quasars we work with are often significantly more noisy, as shown in Figure~\ref{fig:image-examples}. 
To further complicate matters, lensed quasars are partially absent from surveys such as Gaia. Despite Gaia's exceptional resolving power of 180 mas, some quasar images are absent from current data releases because they are too close together, resulting in poorer astrometry which does not pass the filters selecting the sources entering the catalogue. Thus, there is likely a subset of lensed quasars which are absent from surveys like Gaia \citep[even after exploiting subtle signatures, e.g. ][]{Mannucci2022}, despite having been observed.

An image classifier would alleviate this issue and provide an independent model which can improve existing quasar detection methods. Unfortunately, identifying lensed quasars purely from an image is not an unambiguous task. Regardless of the classifier, there will always be sections of the sky where stars masquerade as quasars in a plausible configurations or two separate quasars are coincidentally present on either side of a galaxy, leading to misclassifications. Even panels of expert astronomers have low success rates on this problem: typically before an observation is performed senior astronomers will examine an image of the lensed quasar candidate to judge whether it is sufficiently promising to observe. The success rate of the resulting lensed quasar observations is in the range of only 5--30\%, depending on the collaboration. This low accuracy makes clear the goal of our classifier: not to achieve 100\% accuracy, but to perform image classification competitive with expert astronomers on a large scale. A classifier capable of such performance would be able to parse the millions of known quasars to find the best lensed quasar candidates, something well beyond the reach of a panel of expert astronomers. These better candidates would enable a higher success rate when observing, allowing for more lensed quasar discoveries and a more efficient use of telescope time.

The final problem is that undiscovered lensed quasars likely have different properties to the discovered population of lensed quasars, or else they would have been discovered in previous searches. Concretely, undiscovered lensed quasars likely have a smaller separation, have an image which is reddened or have an image disguised in some other fashion. This adds another challenge to the classification problem: ideally we need a model which can generalise from our small training set to select candidates which are from a different distribution than discovered lensed quasars. This need to generalise breaks assumptions which underpin most machine learning techniques (e.g. that the data are independent and identically distributed across the training and unseen data).

\section{Data}
\subsection{Lensed Quasars}
\label{sec:lensed-quasars}
Our list of known lensed quasars comes from a catalogue by Krone-Martins (priv. comm., 2021). This catalogue is a collation of published lensed quasars as well some some as yet unpublished lensed quasars identified by GraL collaboration (Ducourant, submitted for publication). In this catalogue the coordinates for each image are listed separately. To avoid cluttering the data set with very similar images, one image from each lensed quasar was randomly selected to be included and the rest were discarded. 

Additionally, we compiled old observing logs to create a list of previously observed quasars which were not found to be lenses. These objects were then individually examined and those which looked convincingly like lensed quasars were collated into another list. Although this list contains objects which are not lensed quasars, it does contain objects which fooled a panel of expert astronomers into expending observing resources on them and, given the scarcity of lensed quasars, this was judged (and was demonstrated in early testing) to be sufficiently valuable as to increase classifier performance when included in the set of ``lensed quasars''. 

As an extension to this, we also removed (by hand) the \emph{true} lensed quasars which did not appear to be lensed quasars from the image. As discussed in Section~\ref{sec:difficulties}, there are lensed quasars which do not appear to be likely candidates due to the reddening of one image or a strange configuration. This filtering was done so as to simplify the task facing the prospective classifier in light of the small quantities of labelled data. 

These changes reposition the problem from a classification task seeking to identify lensed quasars, to one seeking to identify objects that plausibly match the appearance of lensed quasars. This is a fine distinction which does not affect the usability of the model, since the population which is now excluded, lensed quasars which do not appear to be lenses, are typically rejected by the astronomers performing the observations.

\subsection{General Population Quasars}
\label{sec:quasars}
Quasars were included from a combination of Milliquas catalogues: 7.7c, 7.7, 7.5b, 7.1b, 6.4 \citep{Milliquas6.4, Milliquas7.2}. These catalogues were first checked for entries within 10~arcsec, randomly removing one entry if a neighbour is found. The catalogues were then checked against more recently published catalogues and any overlap within 10~arcsec led to the removal of the entry from the earlier catalogue. Finally, any entry which was within 10~arcsec of a lensed quasar was removed to avoid images where a lensed quasar appears in the corner of another image. 

It should be noted that although we state the contents of these images are quasars, in reality they are unlabelled data and will rarely contain lensed quasars. To form the labelled set of unlensed quasars we hand classified $\sim$1,000 randomly chosen images from each of the Pan-STARRS survey images and the DESI survey images. The rest of the images form the unlabelled set of data.

\subsection{Pan-STARRS}

In this work we draw images from two surveys, the first of which is the Pan-STARRS1 survey DR2 \citep{PanstarrsProcessing2020,PanstarrsImageProcessing2020,PanstarrsDatabase2020}. The Pan-STARRS survey faces the northern sky, imaging north of $-30^{\circ}$ of declination in five broadband filters: \textit{g}, \textit{r}, \textit{i}, \textit{z}, \textit{y}. Pan-STARRS images are often noisy, as shown in Figure~\ref{fig:image-examples}, but in general do provide \textit{g}, \textit{r} and \textit{i} bands. The images can be downloaded from the University of Hawaii via their Image Cutout Service. 

Unfortunately, downloading data from the servers as FITS files is quite slow with the download time increasing linearly with the number of bands requested. However, in the time taken to download a single band, a 3-colour PNG or JPEG image can be downloaded, allowing us to effectively triple the download rate (note that recent upgrades to the hosting of the Pan-STARRS data have rendered this workaround unnecessary). We experimented with PNG and JPEG formats and found that the JPEG compression (see Appendix~\ref{app:JPEG} for details) enhanced model performance by smoothing the high-frequency noise in the images. We downloaded 3.8 million 64x64 pixel JPEG images, corresponding to 16x16~arcseconds of sky, in the \textit{g}, \textit{r} and \textit{i} bands around the quasars and lensed quasars described in Sections~\ref{sec:lensed-quasars} and~\ref{sec:quasars}.

\subsection{DESI Legacy Surveys}
Due to its location in Hawaii, USA, the Pan-STARRS survey primarily images the northern sky, so we require another survey to image quasars in the southern sky. The DESI Legacy Imaging Survey DR10 (henceforth DESI survey) has good coverage south of $-20^{\circ}$ of declination, but often only in select bands. To match the Pan-STARRS images we again chose 64x64 JPEG images in the \textit{g}, \textit{r} and \textit{i} bands, this time numbering 2 million fields. Unfortunately, the DESI survey has not achieved full coverage in the \textit{g}, \textit{r} and \textit{i} bands, so some of our DESI images are 2-colour.  In order to handle these null values, we opted to fill the values with zeros and pass a flag to the classifiers to indicate in which colour bands null values existed in the original image.
Unlike the Pan-STARRS data, DESI survey data can be quickly downloaded for several channels, allowing us to download FITS files for each band and create the JPEGs locally. We found this resulted in less blurred JPEG images compared to JPEGs downloaded directly from the DESI survey servers.

\begin{figure*}
    \centering
    \includegraphics[width=\textwidth]{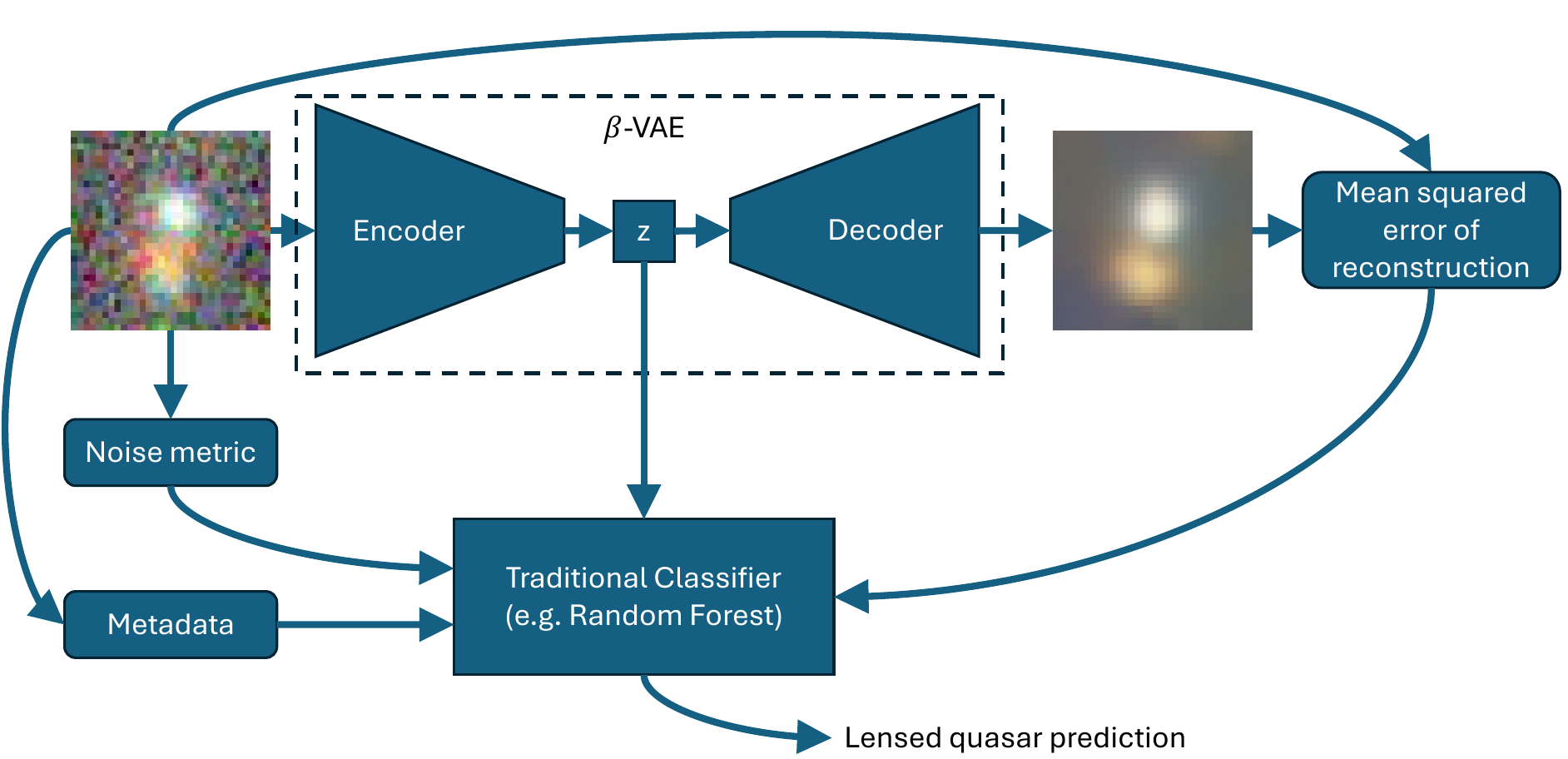}
   \caption{Diagram showing the architecture for the autoencoder-classifier model where z depicts that latent space of the autoencoder. The autoencoder architecture and training is described in Sections~\ref{sec:autoencoder} and \ref{sec:b-VAE}. The noise metric is described in Section~\ref{sec:noise-metric}. The traditional classifiers and metadata are described in Section~\ref{sec:traditional-classifiers}.}
    \label{fig:autoencoder-classifer-diagram}
\end{figure*}

\section{Methods}

Machine learning is finding ever increasing use in astronomy. In the modern era of huge numbers of real time science alerts, machine learning is one of a few tools capable of triaging the deluge of astronomical data collected by our telescopes. Particularly relevant to this work are the studies performed by data brokers such as ALeRCE \citep{Alerce2021} to classify low resolution data. Typically these models are trained on moderate sets of labelled data to classify the stream of data and dispatch science alerts to interested astronomers \citep{Carrasco2021, Sanchez2021a} or to identify anomalies in the data \citep{Sanchez2021b, Perez2023}. They have developed a range of novel techniques, for example, segmenting an image and providing it to a model at different scales \citep{Forster2022}. These studies provided us with inspiration for our model architectures and we hope the present work will in turn spark further new avenues in future.

\subsection{Semi-supervised Learning}

Sitting between supervised learning, where all the data are labelled, and unsupervised learning, where none of the data are labelled, semi-supervised learning is a subset of machine learning consisting of techniques which utilise both labelled and unlabelled data. Typically in a semi-supervised learning problem there exists a large quantity of unlabelled data and a small quantity of labelled data. The aim is to make use of the unlabelled data to achieve better results than could be achieved using the labelled data alone.

\begin{figure*}
    \centering
    \includegraphics[width=\textwidth]{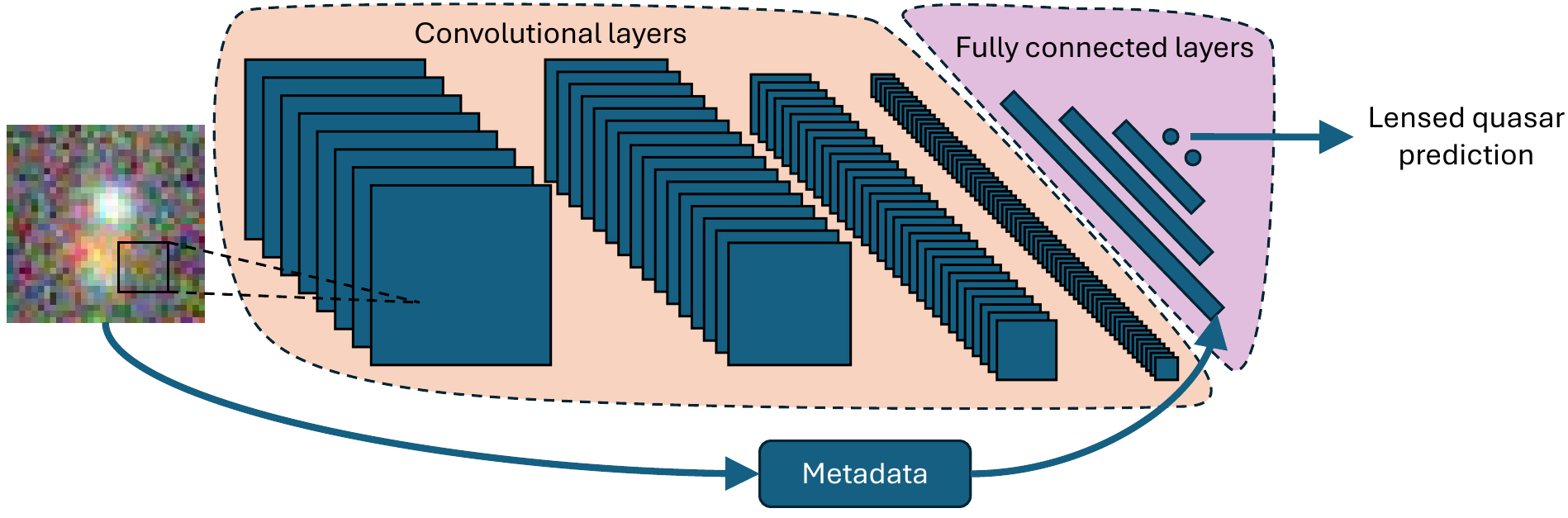}
   \caption{Diagram showing the architecture for the VAT model, as described in Section~\ref{sec:VAT}.}
    \label{fig:VAT-diagram}
\end{figure*}

Identifying lensed quasars is an excellent example of a problem well suited to semi-supervised methods: the number of known lensed quasars is small, but the total population of unlabelled data (e.g. the Milliquas catalogue) is comparatively large. In light of this, we implement two semi-supervised learning methods. The first is an autoencoder-classifier model in which an autoencoder is trained and the outputs of the encoder component are used as inputs to a traditional classifier, described in Sections~\ref{sec:autoencoder}--\ref{sec:classification} and pictured in Figure~\ref{fig:autoencoder-classifer-diagram}. The second approach is to train a convolutional neural network \citep{LeCun1989} using Virtual Adversarial Training \citep[VAT;][hereafter VAT model]{VAT} which regularises the convolutional neural network by providing a penalty for points near the decision boundary, described in Section~\ref{sec:VAT} and pictured in Figure~\ref{fig:VAT-diagram}.

To train these models we split the available data into three sets: training (60\%), validation (20\%) and test (20\%). This division applies to the labelled and unlabelled data and are the sets the VAT model was trained on. For the autoencoder-classifier model we split each set into a labelled and unlabelled component, and provide them to the classifier and autoencoder respectively. Thus the unlabelled component of the training set becomes the training set for the autoencoder and the labelled component of the training set becomes the training set for the classifier.

\subsection{Convolutional Autoencoder}
\label{sec:autoencoder}

\begin{figure}
    \centering
    \includegraphics[width=0.4\columnwidth]{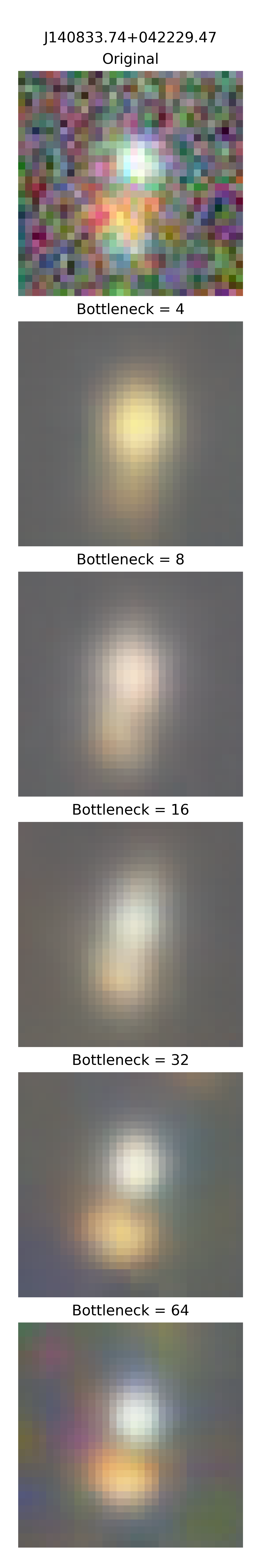}
    \vline
    \includegraphics[width=0.4\columnwidth]{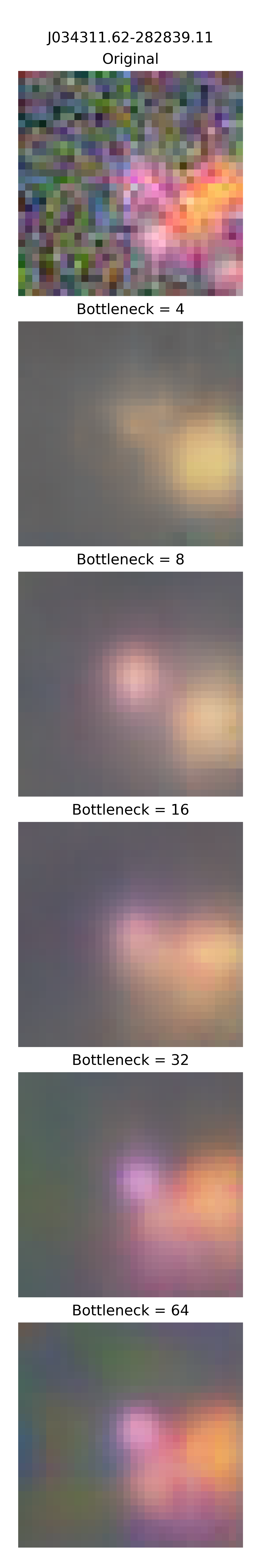}
    \caption{Reconstructions of two (previously unseen by the model) lensed quasars generated by autoencoders with varying bottlenecks, including the best performing autoencoder (described at the end of Section~\ref{sec:classification}). Note that as the dimensionality of the latent space increases the reconstruction fidelity improves, however this often does not translate to increased performance for the classifier as the increased dimensionality of the latent space can reduce performance. These images represent 8x8 arcsec of sky.}
    \label{fig:autoencoder-reconstructions}
\end{figure}

Autoencoders \citep{Autoencoder} are deep neural networks with an hourglass structure which are trained to recreate the input, in our case an image, with their output layer. The initial layer, with the same dimensions as the image, is followed by successively smaller layers until a ``bottleneck'' layer is reached --- this part of the network is the \textit{encoder}. After the bottleneck layer the structure is reversed, with each successive layer getting larger until the original dimensionality is reached --- called the \textit{decoder}. The bottleneck layer typically has small dimensionality (of order 10) which forces the network to reduce the information content of the image into a few values before reconstituting the image. 
Reducing the dimensionality of the bottleneck layer allows less information to be passed to the decoder, resulting in a coarser reconstruction (as shown in Figure~\ref{fig:autoencoder-reconstructions}), but a more information dense latent space.
By training a network in this way we obtain an encoder which is capable of reducing the dimensionality of the input image so that the classification task is tractable for a traditional classifier \citep[e.g. a random forest;][]{RandomForest}.

Constructing an autoencoder to use convolutional layers is a little more complicated than using densely connected layers, but is still a well-established architecture. For our model, the encoder layers are 2D convolutional layers with kernel sizes of 3, strides of 2 and padding of 1 so that the 2D dimensions decrease with successive layers. The first hidden layer has 8 channels and then the number of channels double for each following layer. We found use of the central 32x32 image pixels (8x8 arcseconds) provided the best compromise between maximising context in the image and minimises the number of pixels in the image (and hence the number of weights in the autoencoder). After 4 convolutional layers the network has a densely connected layer of 128 neurons and then the bottleneck layer which varied in size from 4--64 neurons. This process was then reversed to compose the decoder, with the convolutional layers being transposed convolutional layers \citep{Zeiler2010} to reverse the convolutions performed in the encoder.

\subsection{$\beta$-Variational Autoencoder}
\label{sec:b-VAE}
We also trained $\beta$-variational autoencoder models \citep{Higgins2017} which performed better, i.e. resulted in superior classifications compared to the simple autoencoder models. Variational autoencoders \citep[VAEs;][]{Kingma2013} have the same underlying structure as autoencoder models but with the addition of two major elements. The first is that instead of connecting the encoder and decoder directly, the autoencoder outputs a mean and standard deviation for each of the usual bottleneck neurons. The decoder then samples from a normal distribution using the means and standard deviations provided and proceeds with these sampled values. The second difference is that the loss function, given in Equation~\ref{eq:loss-function}, has an added term which penalises the network based on the Kullback-Leibler (KL) divergence \citep{Kullback1951} between the normal distributions described by the means and standard deviations and normal distributions with 0 means and standard deviations of 1. These modifications to the autoencoder encourages it to map images into a latent space at the bottleneck layer which is more meaningful. The random sampling at the bottleneck layer makes it much more difficult for the autoencoder to ``memorise'' training examples by mapping them to specific locations in the latent space, while the added penalty to the loss function ensures the means and standard deviations themselves are hindered from overfitting. Thus, the loss function, $\mathcal{L}(y, \hat{y})$, is:

\begin{align}
    \mathcal{L}(y, \hat{y}) &= \text{MSE}(y, \hat{y}) + \beta \cdot D_{\text{KL}}\left(\mathcal{N}(0, 1) \,||\, \mathcal{N}(\mu, \sigma^2)\right) \label{eq:loss-function}
\end{align}

Where $\text{MSE}(y, \hat{y})$ is the mean squared error between the images and their reconstructions by the model and $D_{\text{KL}}\left(\mathcal{N}(0, 1) \,||\, \mathcal{N}(\mu, \sigma^2)\right)$ is the KL divergence between the means and standard deviations produced by the autoencoder and the standard normal distribution.

The $\beta$ of the $\beta$-VAE refers to the addition of a hyperparameter to weight the contribution of the KL divergence within the loss function. By multiplying the KL divergence by $\beta$, a tuneable hyperparameter, we can set the allowed freedom the model has to customise its latent space, which further increases model performance. Note that a $\beta$ value of 1 reduces to an ``unweighted'' VAE and a value of 0 turns off the penalty altogether. 

All autoencoder models were trained with bottleneck layer sizes ranging from 4--64 (with VAEs this was 4--64 means and 4--64 standard deviations so that the decoder sampled 4--64 values) using a mean squared error loss. Training continued until the loss on the validation set failed to decrease for 5 consecutive epochs.

\subsection{Classification}
\label{sec:classification}

\subsubsection{Parameterising Noise}
\label{sec:noise-metric}

The classifiers were provided with the values from the encoder for each image and the MSE between the image and the autoencoder's reconstruction (hereafter, reconstruction error). The reconstruction error acts as a proxy for the amount of information in the image that the autoencoder has failed to encode in the latent space (and thus failed to recreate with the decoder). This is a useful metric for identifying lensed quasars as images of lensed systems have more information content which the autoencoder struggles to encode in the latent space. However, a great source of confusion for genuine loss of information is the noise present in the original image. If the noise is significant then the reconstruction error can be very large even if the quasar is perfectly reconstructed. Thus, we needed a metric describing the noise level of an image to pass to the classifier, so that it is able to calibrate the reconstruction error based on the noise in the original image.

\begin{figure}
    \centering
    \includegraphics[width=\columnwidth]{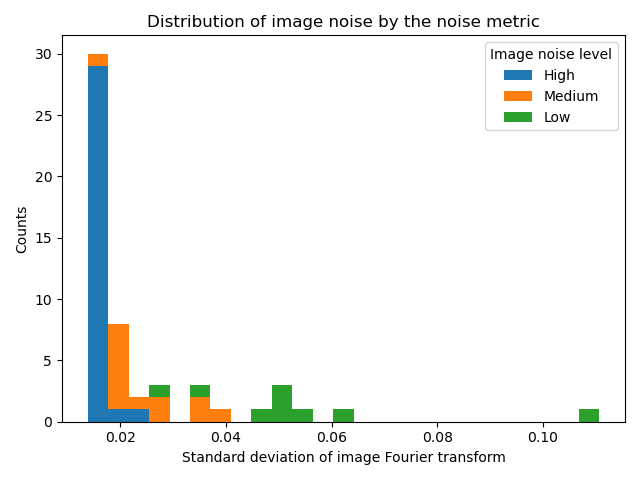}
    \caption{Histogram illustrating the success of our noise metric in categorising images with varying levels of noise. Data plotted consists of $\sim$50 images classified by hand as having low, medium or high noise (colour coded) and are plotted here against the standard deviation of their 2D Fourier transform. The hand-classified low, medium and high noise images are seen to be well separated by our noise metric.}
    \label{fig:FFT-noise}
\end{figure}

As shown in Figure~\ref{fig:image-examples}, the noise in our images is predominantly individual pixels erroneously brightened in one of the colour channels. To capture this phenomena we took the 2D Fourier transform of the images, took the norm of each value and then calculated the standard deviation. Images with the noise present have a lower Fourier standard deviation as the high-frequency noise balances the low-frequency signals (i.e. the quasars) present in the image. By contrast, images without noise lack the high-frequency information and so the low-frequency information dominates the Fourier space, leading to a larger standard deviation due to the large low-frequency values and the small high-frequency values. The success of this metric in splitting the images based on their noisiness is shown in Figure~\ref{fig:FFT-noise}.

\subsubsection{Traditional Classifiers}
\label{sec:traditional-classifiers}

We trained a number of traditional models to classify whether objects were lensed quasars. These classifiers were provided with the latent space values of the chosen autoencoder, the reconstruction errors, the noise metrics, whether each image was from the Pan-STARRS or DESI catalogue and if from the DESI catalogue whether the \textit{g}, \textit{r} and \textit{i} channels were null. For every autoencoder model, 5-fold cross validation tuning was performed to identify the best set of hyperparameters for each classifier. This resulted in classifiers trained for autoencoder and $\beta$-VAE models with bottlenecks of 4/8/16/32/64 and, in the case of the $\beta$-VAE models, various values of $\beta$. The classifiers we chose to train were: random forest \citep{RandomForest}, isolation forest \citep{IsolationForest}, logistic regression \citep{LogisticRegression}, gradient boosting \citep{GradientBoosting}, support vector machine \citep[SVM;][]{SVM}, kernel-SVM \citep{K-SVM}, Gaussian process \citep{GP} and a densely connected artificial neural network. 

\begin{figure}
    \centering
    \includegraphics[width=\columnwidth]{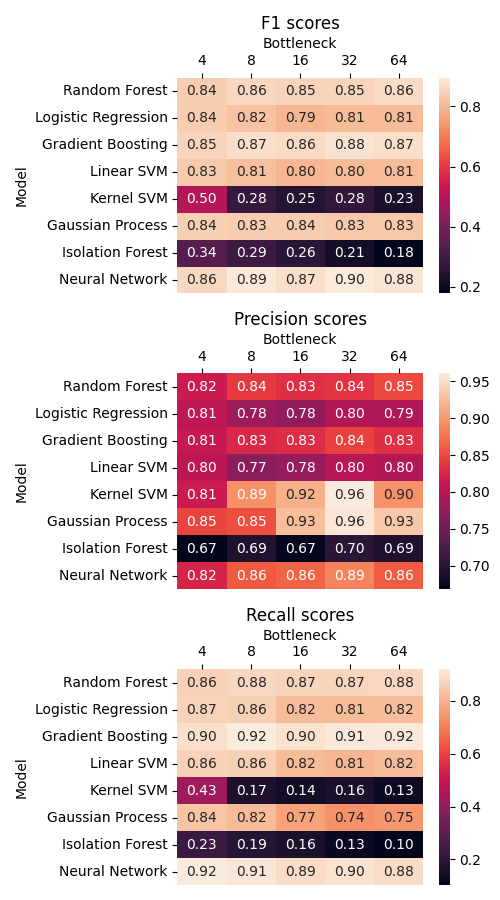}
    \caption{Performance of the traditional classifiers on $\beta$-VAE models with $\beta = 0.0001$ for various bottlenecks. The highest F1 score was achieved by an autoencoder-classifier combination employing a $\beta$-VAE model with $\beta = 0.0001$ and bottleneck $= 8$ and the gradient boosting classifier.}
    \label{fig:classifier-scores}
\end{figure}

The 5-fold cross validation tuning for all models except the neural network comprised randomly searching a predefined space of hyperparameters using scikit-learn's {\tt RandomizedSearchCV} and selecting the set of hyperparameters which achieved the highest F1 score (defined in Appendix~\ref{app:scores}). The neural network was tuned using scikit-learn's {\tt StratifiedKFold} and a grid search over the hyperparameters. The details of the cross validation tuning of each classifier are in Appendix~\ref{app:classifiers}. All models and the random search were seeded with a value of 0, so the results may be recreated. Finally, models were compared using stratified 5-fold cross validation using {\tt StratifiedKFold} with the median F1, precision and recall scores shown in Figure~\ref{fig:classifier-scores}

The best autoencoder-classifer combination achieved an F1 score of 0.90 by using a $\beta$-VAE with $\beta = 0.0001$ (larger and smaller $\beta$ values were tested) and a bottleneck of 32 combined with a densely connected artificial neural network with two hidden layers of 32 nodes and a weight decay of 0.003. The performance of tuned classifiers combined with $\beta$-VAEs with $\beta = 0.0001$ are shown in Figure~\ref{fig:classifier-scores}.

\subsection{Virtual Adversarial Training}
\label{sec:VAT}

VAT is a technique designed to enhance the robustness and accuracy of deep learning models in tasks where labelled data are scarce but unlabelled data are abundant. Readers are referred to \citet{VAT} for full details, but we give a brief overview here. At its core, VAT operates by augmenting the labelled data by classifying the unlabelled data and then penalising the model if the classifications change after a small adversarial perturbation to the original image. To achieve this, the loss function is composed of two terms. The first is the normal loss associated with the chosen deep classifier, in our case cross entropy loss, together with any regularisation. The second term is calculated by classifying the unlabelled images, making a small adversarial perturbation and then classifying the perturbed images. The number of images, multiplied by a tuneable hyperparameter, which change their classification under this perturbation becomes the second loss term for the model. Adversarially perturbing the image with a deep model is straightforward as the derivative of the image can be taken with respect to the final classification to give a vector pointing in the adversarial direction. The image is then modified by adding this vector to the image to create a new image which has been modified in the worst possible way for the classifier. This new image can then simply be passed through the model to result in another classification to be compared to the old classification.

For our purposes we highlight a few key details. Firstly, this added penalty to the loss function allows the model to learn from the numerous unlabelled data available. Penalising the model based on the reclassification of the unlabelled data under the adversarial perturbation encourages our model to shift its decision boundary to low-density areas of the image space. This in turn leads to fewer points changing their classification when perturbed, resulting in a lower penalty for the loss function. This behaviour is ideal since we seek a boundary which sits between images depicting lensed quasars and images that do not. 

VAT also allows us to train an end-to-end classifier to identify lensed quasars --- as opposed to the autoencoder-classifier combination discussed previously. This has a number of advantages. Firstly, by using a single model, the information imparted by the latent features does not have to be relearnt. In the autoencoder-classifier structure, the classifier has to learn (with very few training examples) the meaning of the bottleneck features passed to it from the autoencoder. However, since the VAT model sees the process from start to end, there is no need for it to relearn any information about the latent features. Furthermore, because the VAT model is not trained only on the labelled data (like the classifier in the autoencoder-classifier structure), it has some exposure to the unlabelled data points and how they might be classified, driving the algorithm to better out-of-distribution performance.

Another advantage is that the training process is easier for the VAT model than the autoencoder-classifier combination. When training the VAT model we are able to use the entirety of the training set to train the model, instead of splitting the set into a fraction for the autoencoder and a fraction for the classifier. Hyperparameter tuning is similarly straightforward as the VAT model can simply be trained and evaluated for each set of hyperparameters, as opposed to training the autoencoder, then generating the bottleneck features for the classifier and then training the classifier for evaluation. 

Finally, because of the end-to-end structure, the VAT model is able to adjust the latent space to optimise the classification. By contrast, the two model structure of the autoencoder-classifier has the autoencoder optimising the bottleneck latent space for image reconstruction, not classification. This difference allows the VAT model to focus primarily on features which aid classification, e.g. whether any of the sources are extended (useful for differentiating galaxies from quasars), rather than their precise shape (a requirement for image reconstruction).

In this work we train a convolutional neural network using VAT. We structure it similarly to the autoencoder described in Section~\ref{sec:autoencoder}, starting three channels in the input layer and then having four convolutional layers with 8, 16, 32, and 64 channels. Each layer has a kernel size of 3, a stride of 2 and a padding of 1. Between each convolutional layer was a batch normalisation \citep{Ioffe2015} layer and a leaky ReLU activation function. We found that the best F1 score was achieved by flattening the network to 260 nodes before passing through 4 densely connected layers which have 256, then 128, 64 and 2 features. Between each of the densely connected layers is a leaky ReLU activation function and a batch normalisation layer, with a sigmoid activation after the final densely connected layer. We experimented with varying values of dropout to regularise the network, but ultimately found that the network performed best without dropout.

\section{Discussion}
\subsection{Comparison}

\begin{figure}
    \centering
    \includegraphics[width=\columnwidth]{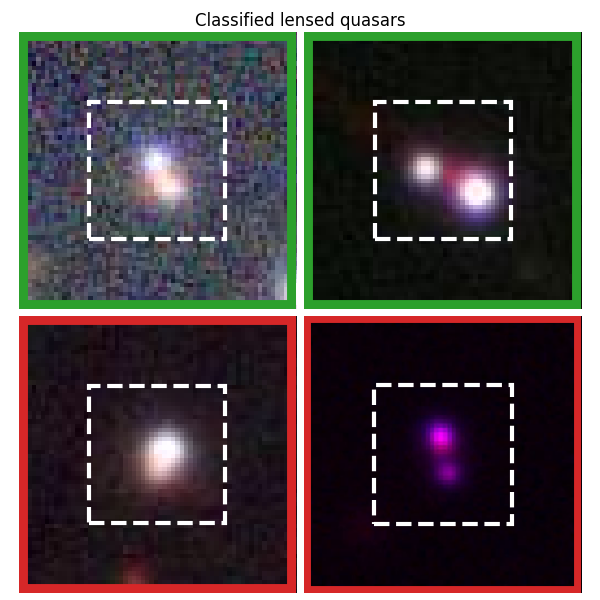}
    \caption{Some example classifications of lensed quasars from the best autoencoder-classifier model. The top row are two correctly classified lensed quasars (as indicated by the green border), while the bottom row are two lensed quasars which the model fails to identify (as indicated by the red border). In each image the white dashed square indicates the inner 32x32 pixels (8x8 arcseconds) which were passed to the autoencoder. In the bottom left image the model likely misclassifies the lens as the reddening caused by the lensing galaxy affects one image significantly more than the other. In the bottom right image the model may be confused by the colouring induced by the $i$ band (blue channel) being null in this area of the sky in the DESI survey.}
    \label{fig:classifications}
\end{figure}

The best performing autoencoder-classifier model achieved an F1 score of 0.897 on the test set, significantly outperforming the best VAT model with an F1 score of 0.58. Thus, on our relatively clean data set the simplicity of the traditional classifier outperforms the flexibility of the convolutional neural network trained via VAT. Some example classifications from the autoencoder-classifier model are shown in Figure~\ref{fig:classifications}.

However, to generate lensed quasar candidates for follow-up observation, the models were presented with unlabelled data from the test set and tasked to rank images from most likely to contain a lensed quasar to least likely. There are many more unlabelled images than the original training data so this unlabelled data contained kinds of images that the classifiers have not seen before. While both classifiers generate some excellent candidates, anecdotally they perform very similarly on this task. Unfortunately when confronted with images of crowded stellar fields, the autoencoder-classifier model was found to rank these highly despite an extremely low likelihood for the presence of lensed quasars. This misclassification is understandable as, by coincidence, these images contain asterisms with approximately the correct orientation and colour as lensed quasars while none of the labelled training set (upon which the classifier was trained) contained similar images. By contrast, the VAT model did not suffer similar difficulties with crowded stellar fields, likely because the end-to-end approach exposes the model to crowded stellar fields in the unlabelled training data, leading the model to correctly identify that they are poor lensed quasar candidates. Despite the end-to-end approach, the VAT model does still falsely identify interacting galaxies as lensed quasar candidates. Typically the interacting galaxies it misclassifies have a quasar in each galaxy and then the interacting galaxies between them: a configuration bearing a strong superficial resemblance to lensed quasars.

\subsection{On-sky Results}

% https://docs.google.com/spreadsheets/d/1-OY5xSzoprhIoDp4DjebDSTCUlRY4Aj5WxlG8yVlgeQ/edit?usp=sharing
% https://docs.google.com/spreadsheets/d/1AH3as22XH0FIdz43LCRzDs3LDy6NhQwvLoAmsJVSMac/edit?usp=sharing

The candidates generated by the models were vetted by a team of expert astronomers in the GraL collaboration from which a handful were selected for observation. We obtained optical spectroscopic follow-up of five candidates using the Low-Resolution Imaging Spectrometer \citep[LRIS; ][]{Oke1995} at the W.M. Keck Observatory between January and March 2023. Table~1 provides information about the observed sources.  For all targets, we observed using the $1\farcs5$ slit, the 600 line blue grism (blazed at 4000~\AA), the 5600~\AA\, dichroic, and the 400 line red grating (blazed at 8500~\AA). The slits were aligned based on the candidate lens configuration and the data were reduced using standard techniques within IRAF.

\begin{table*}
    \caption{Keck spectroscopy of strong lens candidates. We have omitted the target which we were unable to resolve.}
    \label{tab:spectroscopy}
    \begin{tabular}{l|r|r|r|l}
        \hline
        Name                     & Observation Date (UT) & PA (deg) & Exposure time (s) & Notes  \\
        \hline
        GRALJ121709.65-025621.60 &  2023 Jan 22          &    122   & 2$\times$300      & $z=1.464$ quasar + $z=0.165$ emission-line galaxy \\
        GRALJ140833.73+042229.98 &  2023 Jan 22          &    173   & 2$\times$600      & $z=2.998$ quasar lensed by $z=0.542$ early-type galaxy\\
        GRALJ142216.45-141650.05 &  2023 Mar 28          &      5   &          300      & $z=1.703$ quasar + Galactic star\\
        GRALJ174005.86+221100.98 &  2023 Mar 28          &    249   &          300      & $z=1.405$ quasar + Galactic star\\
    \end{tabular}
\end{table*}

\begin{figure}
    \centering
    \includegraphics[width=0.7\columnwidth]{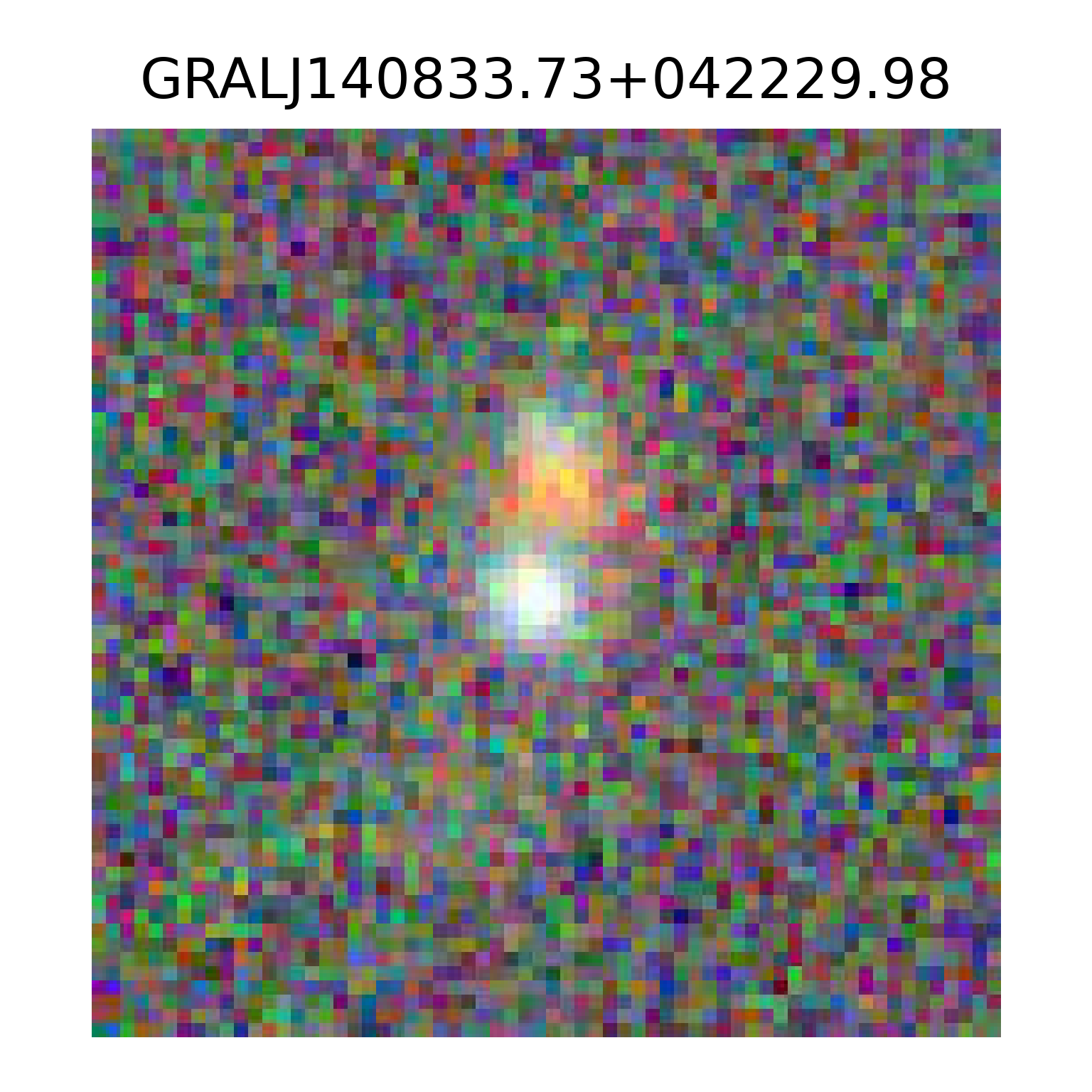}
    \caption{Newly discovered quasar GRALJ140833.73+042229.98, internally named ``the Snowman''.}
    \label{fig:discovered-qso}
\end{figure}

This led to the confirmation of one new lensed quasar, GRALJ140833.73+042229.98, for which we obtained spectra of the lensed $z=2.998$ quasar flanking the early-type galaxy lens at $z=0.542$ (Figure~\ref{fig:discovered-qso}). A second candidate was unable to be resolved. The remaining candidates all contain asterisms of a quasar accompanied by a foreground source (i.e., a Galactic star in two cases and an emission-line galaxy in the third case).

The observations were performed in parallel with the development of these models, so the successful detection came from a previous iteration of the autoencoder-classifier model, one which used a simple convolutional autoencoder and a random forest classifier. The selected candidates were a combination of candidates from both
models, but by coincidence the observed candidates were all from the autoencoder-classifier model.

This success rate, one strong confirmation and one unresolved target from five observations, is competitive with current state-of-the-art techniques. For example, compared to the quantum annealing approach used by the GraL collaboration (Krone-Martins priv. comm., 2024), our approach has the advantage of using only a GPU or CPU as opposed to the quantum computer required for the quantum annealing. Additionally, since we use distinct data (the models in this paper use image data, while the quantum annealing approach uses a mix of photometric and astrometric data) similar to that used in \citet{Krone-Martins2018}, improved performance could be achieved by combining the predictions made by the two classifiers.

\subsection{Model Improvements}

% While the autoencoder-classifier model achieves the best F1 score on our test data, we believe there is limited room for improvement to this approach, due to the inherent training objective misalignment between the autoencoder and the classifier, described in Section~\ref{sec:VAT}. 
Increasing the quantity of labelled data would likely improve the performance of both the autoencoder-classifier model and the VAT model. Currently, the performance of both models suffer when presented with images which are unlike images contained in the labelled training set. 
To diversify the training set we tried including unlabelled quasars as unlensed quasars. However, this did not increase performance of the autoencoder-classifier model and degraded performance of the VAT model, presumably due to the accidental inclusion of mislabelled lensed quasars into the training set.
While significantly increasing the number of lensed quasar images is infeasible on short timescales, increasing the training set by labelling unlensed quasars would increase performance, particularly if the labelling was targeted to examples where the models struggled. 
We estimate that if 10,000 more images were labelled, this would significantly increase the diversity of images in the labelled training data and would improve model performance in these unusual cases. 

A large number of unlensed labels could be automatically applied by cross-matching the images with a catalogue like Gaia. If Gaia detects all of the sources visible in the image with a low renormalised unit weight error (a measure of the reliability of Gaia's astrometric fit for a source) then images could be labelled as unlensed quasars if the configuration was unlikely for a lens or any of the objects had significant proper motion/parallax. Labelling the data in this way would likely improve performance for our most common source of confusion: images which contain a quasar and a similarly coloured star. 

Although significantly increasing the number of confirmed lensed quasars in the training data is infeasible using the approach described in this work, progress could be achieved by incorporating simulated images into the training data. This was done by \citet{Akhazhanov2022}, which produced significantly larger labelled sets. However, it introduces challenges with corrupting the simulated images to match the Pan-STARRS/DESI survey images. Another difficulty is that by training on lensed quasars from simulations, we bias the model toward detecting objects like those that we generate, so care has to be taken to generate a sufficiently diverse range of lensed quasar simulations.

Including other spectral bands in the images would also likely increase model performance. Quasar spectra are distinct from stellar spectra, so providing greater spectral information to the classifier in the form of additional bands should increase performance. In our case we chose to use only the \textit{g}, \textit{r} and \textit{i} bands due to time limitations in downloading the data from the  Pan-STARRS survey. However, with more time or more direct access to the \textit{z}-band data (which also exists in the DESI survey), this could be included. 

The model predictions could also be improved by taking advantage of the area of the sky in which both the Pan-STARRS and DESI surveys have overlapping coverage. A simple approach might be to simply average the predictions of the Pan-STARRS and DESI images. However, this only affects predictions in the shared area of sky. A more sophisticated approach would be to provide an added regularisation term to the VAT model which penalises it for any difference in prediction between a Pan-STARRS image and a DESI image in cases where both images exist. By taking this approach, the model is encouraged to learn a stronger correspondence between the two surveys which should improve performance across the entire sky.

\section{Conclusions}

In this paper we explore the prospect of using semi-supervised learning to identify lensed quasars from images drawn from the Pan-STARRS survey and the DESI survey. This classification task presents interesting challenges. Firstly, the number of known lensed quasars sits at 650: a tiny sample for modern machine learning tasks. Secondly, the images to which we have access to are often noisy, and the images available in the northern sky are collected from a different telescope than the images of the southern sky and thus exhibit different characteristics and systematics. Thirdly, identifying lensed quasars from images alone is considered an ambiguous task; teams of expert astronomers typically achieve success rates in the 5--30\% range. Finally, unidentified lensed quasars likely have different characteristics compared to the found lensed quasars; most likely they have smaller separations and/or one of the quasar images is obscured. Differences between the identified lensed quasars (our training data) and the target unidentified lensed quasar population breaks the assumptions that underpin most machine learning approaches.

Despite these challenges, we develop two models which are capable of distinguishing between lensed and unlensed quasars from image data. The first approach is to train an $\beta$-VAE to recreate images of quasars, of which we source a few million. The labelled data is passed through the $\beta$-VAE and the latent space is passed to a classifier, the best performing of which was found to be a densely connected artificial neural networkgradient boosting. The second approach is to train a convolutional neural network using VAT to allow the network to be trained on both labelled and unlabelled data, thus taking advantage of the significant number of quasars to increase the performance of the network despite our small set of labelled data. Of these two approaches we find that the autoencoder-classifier model achieves a better F1 score, 0.897, on our clean dataset compared to our VAT model, 0.58. However, when the models are tasked with finding undiscovered lensed quasars they perform with roughly equal success as the end-to-end approach of the VAT model allows it to generalise well. Both models still struggle with images which have few similarities to images in the labelled set (e.g., crowded fields). We conducted observations at the W.M. Keck Observatory of 5 lensed quasar candidates delivered by our classifiers, confirming one to be a previously undiscovered lensed quasar GRALJ140833.73+042229.98, three others to be interlopers consisting of a quasar and a nearby star and one candidate an unconfirmed lens requiring further follow-up.

We chose to train classifiers to identify lensed quasars based on image data alone despite the challenges involved. Their demonstrated utility holds out the promise that, when combined with existing methods, e.g. quantum annealing, \citet{Krone-Martins2018} or \citet{Akhazhanov2022}, significant increases in performance may result. Particularly since the quantum annealing or \citet{Krone-Martins2018} does not use image data, this work should act as an independent model which, when combined with these approaches should lead to better performance than either could achieve alone.

With increasingly powerful all-sky surveys such as Gaia, DESI, Euclid \citep{Euclid} and the forthcoming Legacy Survey of Space and Time (LSST; expected to generate 20~TB of data \textit{per night}) machine learning will become an increasingly core part of the modern astronomer's toolkit. In this work we demonstrate using semi-supervised learning to take advantage of these large sets of data. The approaches we outline here can be applied to any problem seeking to extract signals of rare events or features from colossal surveys. Using semi-supervised learning to parse the torrent of data delivered by modern surveys will populate classes of objects at previously unthinkable rates --- an exciting opportunity for both machine learning and astronomy.

\section*{Acknowledgements}

The work of D. Stern was carried out at the Jet Propulsion Laboratory, California Institute of Technology, under a contract with the National Aeronautics and Space Administration (80NM0018D0004).

The Pan-STARRS1 Surveys (PS1) and the PS1 public science archive have been made possible through contributions by the Institute for Astronomy, the University of Hawaii, the Pan-STARRS Project Office, the Max-Planck Society and its participating institutes, the Max Planck Institute for Astronomy, Heidelberg and the Max Planck Institute for Extraterrestrial Physics, Garching, The Johns Hopkins University, Durham University, the University of Edinburgh, the Queen's University Belfast, the Harvard-Smithsonian Center for Astrophysics, the Las Cumbres Observatory Global Telescope Network Incorporated, the National Central University of Taiwan, the Space Telescope Science Institute, the National Aeronautics and Space Administration under Grant No. NNX08AR22G issued through the Planetary Science Division of the NASA Science Mission Directorate, the National Science Foundation Grant No. AST-1238877, the University of Maryland, Eotvos Lorand University (ELTE), the Los Alamos National Laboratory, and the Gordon and Betty Moore Foundation.

The Legacy Surveys consist of three individual and complementary projects: the Dark Energy Camera Legacy Survey (DECaLS; Proposal ID \#2014B-0404; PIs: David Schlegel and Arjun Dey), the Beijing-Arizona Sky Survey (BASS; NOAO Prop. ID \#2015A-0801; PIs: Zhou Xu and Xiaohui Fan), and the Mayall z-band Legacy Survey (MzLS; Prop. ID \#2016A-0453; PI: Arjun Dey). DECaLS, BASS and MzLS together include data obtained, respectively, at the Blanco telescope, Cerro Tololo Inter-American Observatory, NSF’s NOIRLab; the Bok telescope, Steward Observatory, University of Arizona; and the Mayall telescope, Kitt Peak National Observatory, NOIRLab. Pipeline processing and analyses of the data were supported by NOIRLab and the Lawrence Berkeley National Laboratory (LBNL). The Legacy Surveys project is honored to be permitted to conduct astronomical research on Iolkam Du’ag (Kitt Peak), a mountain with particular significance to the Tohono O’odham Nation.

NOIRLab is operated by the Association of Universities for Research in Astronomy (AURA) under a cooperative agreement with the National Science Foundation. LBNL is managed by the Regents of the University of California under contract to the U.S. Department of Energy.

This project used data obtained with the Dark Energy Camera (DECam), which was constructed by the Dark Energy Survey (DES) collaboration. Funding for the DES Projects has been provided by the U.S. Department of Energy, the U.S. National Science Foundation, the Ministry of Science and Education of Spain, the Science and Technology Facilities Council of the United Kingdom, the Higher Education Funding Council for England, the National Center for Supercomputing Applications at the University of Illinois at Urbana-Champaign, the Kavli Institute of Cosmological Physics at the University of Chicago, Center for Cosmology and Astro-Particle Physics at the Ohio State University, the Mitchell Institute for Fundamental Physics and Astronomy at Texas A\&M University, Financiadora de Estudos e Projetos, Fundacao Carlos Chagas Filho de Amparo, Financiadora de Estudos e Projetos, Fundacao Carlos Chagas Filho de Amparo a Pesquisa do Estado do Rio de Janeiro, Conselho Nacional de Desenvolvimento Cientifico e Tecnologico and the Ministerio da Ciencia, Tecnologia e Inovacao, the Deutsche Forschungsgemeinschaft and the Collaborating Institutions in the Dark Energy Survey. The Collaborating Institutions are Argonne National Laboratory, the University of California at Santa Cruz, the University of Cambridge, Centro de Investigaciones Energeticas, Medioambientales y Tecnologicas-Madrid, the University of Chicago, University College London, the DES-Brazil Consortium, the University of Edinburgh, the Eidgenossische Technische Hochschule (ETH) Zurich, Fermi National Accelerator Laboratory, the University of Illinois at Urbana-Champaign, the Institut de Ciencies de l’Espai (IEEC/CSIC), the Institut de Fisica d’Altes Energies, Lawrence Berkeley National Laboratory, the Ludwig Maximilians Universitat Munchen and the associated Excellence Cluster Universe, the University of Michigan, NSF’s NOIRLab, the University of Nottingham, the Ohio State University, the University of Pennsylvania, the University of Portsmouth, SLAC National Accelerator Laboratory, Stanford University, the University of Sussex, and Texas A\&M University.

BASS is a key project of the Telescope Access Program (TAP), which has been funded by the National Astronomical Observatories of China, the Chinese Academy of Sciences (the Strategic Priority Research Program “The Emergence of Cosmological Structures” Grant \# XDB09000000), and the Special Fund for Astronomy from the Ministry of Finance. The BASS is also supported by the External Cooperation Program of Chinese Academy of Sciences (Grant \# 114A11KYSB20160057), and Chinese National Natural Science Foundation (Grant \# 12120101003, \# 11433005).

The Legacy Survey team makes use of data products from the Near-Earth Object Wide-field Infrared Survey Explorer (NEOWISE), which is a project of the Jet Propulsion Laboratory/California Institute of Technology. NEOWISE is funded by the National Aeronautics and Space Administration.

The Legacy Surveys imaging of the DESI footprint is supported by the Director, Office of Science, Office of High Energy Physics of the U.S. Department of Energy under Contract No. DE-AC02-05CH1123, by the National Energy Research Scientific Computing Center, a DOE Office of Science User Facility under the same contract; and by the U.S. National Science Foundation, Division of Astronomical Sciences under Contract No. AST-0950945 to NOAO.

In addition to those mentioned in the main text, we would also like to acknowledge other python packages used as part of this work: NumPy \citep{NumPy}, Matplotlib \citep{Matplotlib}, seaborn \citep{seaborn}, pandas \citep{pandas2010, pandas}, Astropy \citep{astropy2013, astropy2018, astropy2022}, astroquery \citep{astroquery}, PyTorch \citep{PyTorch}, scikit-learn \citep{scikit-learn}, Pillow \citep{PIL}, scikit-image \citep{scikit-image}, VAT-pytorch \citep{VATLoss} and ray \citep{ray}.

%%%%%%%%%%%%%%%%%%%%%%%%%%%%%%%%%%%%%%%%%%%%%%%%%%
\section*{Data Availability}

The code underlying this article are available on Zenodo \citep{zenodo-code}. The data may be made available upon personal request.

%%%%%%%%%%%%%%%%%%%% REFERENCES %%%%%%%%%%%%%%%%%%

% The best way to enter references is to use BibTeX:

\bibliographystyle{mnras}
\bibliography{example} % if your bibtex file is called example.bib

@ARTICLE{2023Coryn,
       author = {{Gaia Collaboration} and {Bailer-Jones}, C.~A.~L. and {Teyssier}, D. and {Delchambre}, L. and {Ducourant}, C. and {Garabato}, D. and {Hatzidimitriou}, D. and {Klioner}, S.~A. and {Rimoldini}, L. and {Bellas-Velidis}, I. and {Carballo}, R. and {Carnerero}, M.~I. and {Diener}, C. and {Fouesneau}, M. and {Galluccio}, L. and {Gavras}, P. and {Krone-Martins}, A. and {Raiteri}, C.~M. and {Teixeira}, R. and {Brown}, A.~G.~A. and {Vallenari}, A. and {Prusti}, T. and {de Bruijne}, J.~H.~J. and {Arenou}, F. and {Babusiaux}, C. and {Biermann}, M. and {Creevey}, O.~L. and {Evans}, D.~W. and {Eyer}, L. and {Guerra}, R. and {Hutton}, A. and {Jordi}, C. and {Lammers}, U.~L. and {Lindegren}, L. and {Luri}, X. and {Mignard}, F. and {Panem}, C. and {Pourbaix}, D. and {Randich}, S. and {Sartoretti}, P. and {Soubiran}, C. and {Tanga}, P. and {Walton}, N.~A. and {Bastian}, U. and {Drimmel}, R. and {Jansen}, F. and {Katz}, D. and {Lattanzi}, M.~G. and {van Leeuwen}, F. and {Bakker}, J. and {Cacciari}, C. and {Casta{\~n}eda}, J. and {De Angeli}, F. and {Fabricius}, C. and {Fr{\'e}mat}, Y. and {Guerrier}, A. and {Heiter}, U. and {Masana}, E. and {Messineo}, R. and {Mowlavi}, N. and {Nicolas}, C. and {Nienartowicz}, K. and {Pailler}, F. and {Panuzzo}, P. and {Riclet}, F. and {Roux}, W. and {Seabroke}, G.~M. and {Sordo}, R. and {Th{\'e}venin}, F. and {Gracia-Abril}, G. and {Portell}, J. and {Altmann}, M. and {Andrae}, R. and {Audard}, M. and {Benson}, K. and {Berthier}, J. and {Blomme}, R. and {Burgess}, P.~W. and {Busonero}, D. and {Busso}, G. and {C{\'a}novas}, H. and {Carry}, B. and {Cellino}, A. and {Cheek}, N. and {Clementini}, G. and {Damerdji}, Y. and {Davidson}, M. and {de Teodoro}, P. and {Nu{\~n}ez Campos}, M. and {Dell'Oro}, A. and {Esquej}, P. and {Fern{\'a}ndez-Hern{\'a}ndez}, J. and {Fraile}, E. and {Garc{\'\i}a-Lario}, P. and {Gosset}, E. and {Haigron}, R. and {Halbwachs}, J. -L. and {Hambly}, N.~C. and {Harrison}, D.~L. and {Hern{\'a}ndez}, J. and {Hestroffer}, D. and {Hodgkin}, S.~T. and {Holl}, B. and {Jan{\ss}en}, K. and {Jevardat de Fombelle}, G. and {Jordan}, S. and {Lanzafame}, A.~C. and {L{\"o}ffler}, W. and {Marchal}, O. and {Marrese}, P.~M. and {Moitinho}, A. and {Muinonen}, K. and {Osborne}, P. and {Pancino}, E. and {Pauwels}, T. and {Recio-Blanco}, A. and {Reyl{\'e}}, C. and {Riello}, M. and {Roegiers}, T. and {Rybizki}, J. and {Sarro}, L.~M. and {Siopis}, C. and {Smith}, M. and {Sozzetti}, A. and {Utrilla}, E. and {van Leeuwen}, M. and {Abbas}, U. and {{\'A}brah{\'a}m}, P. and {Abreu Aramburu}, A. and {Aerts}, C. and {Aguado}, J.~J. and {Ajaj}, M. and {Aldea-Montero}, F. and {Altavilla}, G. and {{\'A}lvarez}, M.~A. and {Alves}, J. and {Anderson}, R.~I. and {Anglada Varela}, E. and {Antoja}, T. and {Baines}, D. and {Baker}, S.~G. and {Balaguer-N{\'u}{\~n}ez}, L. and {Balbinot}, E. and {Balog}, Z. and {Barache}, C. and {Barbato}, D. and {Barros}, M. and {Barstow}, M.~A. and {Bartolom{\'e}}, S. and {Bassilana}, J. -L. and {Bauchet}, N. and {Becciani}, U. and {Bellazzini}, M. and {Berihuete}, A. and {Bernet}, M. and {Bertone}, S. and {Bianchi}, L. and {Binnenfeld}, A. and {Blanco-Cuaresma}, S. and {Boch}, T. and {Bombrun}, A. and {Bossini}, D. and {Bouquillon}, S. and {Bragaglia}, A. and {Bramante}, L. and {Breedt}, E. and {Bressan}, A. and {Brouillet}, N. and {Brugaletta}, E. and {Bucciarelli}, B. and {Burlacu}, A. and {Butkevich}, A.~G. and {Buzzi}, R. and {Caffau}, E. and {Cancelliere}, R. and {Cantat-Gaudin}, T. and {Carlucci}, T. and {Carrasco}, J.~M. and {Casamiquela}, L. and {Castellani}, M. and {Castro-Ginard}, A. and {Chaoul}, L. and {Charlot}, P. and {Chemin}, L. and {Chiaramida}, V. and {Chiavassa}, A. and {Chornay}, N. and {Comoretto}, G. and {Contursi}, G. and {Cooper}, W.~J. and {Cornez}, T. and {Cowell}, S. and {Crifo}, F. and {Cropper}, M. and {Crosta}, M. and {Crowley}, C. and {Dafonte}, C. and {Dapergolas}, A. and {David}, P. and {de Laverny}, P. and {De Luise}, F. and {De March}, R. and {De Ridder}, J. and {de Souza}, R. and {de Torres}, A. and {del Peloso}, E.~F. and {del Pozo}, E. and {Delbo}, M. and {Delgado}, A. and {Delisle}, J. -B. and {Demouchy}, C. and {Dharmawardena}, T.~E. and {Diakite}, S. and {Distefano}, E. and {Dolding}, C. and {Enke}, H. and {Fabre}, C. and {Fabrizio}, M. and {Faigler}, S. and {Fedorets}, G. and {Fernique}, P. and {Figueras}, F. and {Fournier}, Y. and {Fouron}, C. and {Fragkoudi}, F. and {Gai}, M. and {Garcia-Gutierrez}, A. and {Garcia-Reinaldos}, M. and {Garc{\'\i}a-Torres}, M. and {Garofalo}, A. and {Gavel}, A. and {Gerlach}, E. and {Geyer}, R. and {Giacobbe}, P. and {Gilmore}, G. and {Girona}, S. and {Giuffrida}, G. and {Gomel}, R. and {Gomez}, A. and {Gonz{\'a}lez-N{\'u}{\~n}ez}, J. and {Gonz{\'a}lez-Santamar{\'\i}a}, I. and {Gonz{\'a}lez-Vidal}, J.~J. and {Granvik}, M. and {Guillout}, P. and {Guiraud}, J. and {Guti{\'e}rrez-S{\'a}nchez}, R. and {Guy}, L.~P. and {Hauser}, M. and {Haywood}, M. and {Helmer}, A. and {Helmi}, A. and {Sarmiento}, M.~H. and {Hidalgo}, S.~L. and {Hilger}, T. and {H{\l}adczuk}, N. and {Hobbs}, D. and {Holland}, G. and {Huckle}, H.~E. and {Jardine}, K. and {Jasniewicz}, G. and {Jean-Antoine Piccolo}, A. and {Jim{\'e}nez-Arranz}, {\'O}. and {Juaristi Campillo}, J. and {Julbe}, F. and {Karbevska}, L. and {Kervella}, P. and {Khanna}, S. and {Kontizas}, M. and {Kordopatis}, G. and {Korn}, A.~J. and {K{\'o}sp{\'a}l}, {\'A}. and {Kostrzewa-Rutkowska}, Z. and {Kruszy{\'n}ska}, K. and {Kun}, M. and {Laizeau}, P. and {Lambert}, S. and {Lanza}, A.~F. and {Lasne}, Y. and {Le Campion}, J. -F. and {Lebreton}, Y. and {Lebzelter}, T. and {Leccia}, S. and {Leclerc}, N. and {Lecoeur-Taibi}, I. and {Liao}, S. and {Licata}, E.~L. and {Lindstr{\o}m}, H.~E.~P. and {Lister}, T.~A. and {Livanou}, E. and {Lobel}, A. and {Lorca}, A. and {Loup}, C. and {Madrero Pardo}, P. and {Magdaleno Romeo}, A. and {Managau}, S. and {Mann}, R.~G. and {Manteiga}, M. and {Marchant}, J.~M. and {Marconi}, M. and {Marcos}, J. and {Marcos Santos}, M.~M.~S. and {Mar{\'\i}n Pina}, D. and {Marinoni}, S. and {Marocco}, F. and {Marshall}, D.~J. and {Martin Polo}, L. and {Mart{\'\i}n-Fleitas}, J.~M. and {Marton}, G. and {Mary}, N. and {Masip}, A. and {Massari}, D. and {Mastrobuono-Battisti}, A. and {Mazeh}, T. and {McMillan}, P.~J. and {Messina}, S. and {Michalik}, D. and {Millar}, N.~R. and {Mints}, A. and {Molina}, D. and {Molinaro}, R. and {Moln{\'a}r}, L. and {Monari}, G. and {Mongui{\'o}}, M. and {Montegriffo}, P. and {Montero}, A. and {Mor}, R. and {Mora}, A. and {Morbidelli}, R. and {Morel}, T. and {Morris}, D. and {Muraveva}, T. and {Murphy}, C.~P. and {Musella}, I. and {Nagy}, Z. and {Noval}, L. and {Oca{\~n}a}, F. and {Ogden}, A. and {Ordenovic}, C. and {Osinde}, J.~O. and {Pagani}, C. and {Pagano}, I. and {Palaversa}, L. and {Palicio}, P.~A. and {Pallas-Quintela}, L. and {Panahi}, A. and {Payne-Wardenaar}, S. and {Pe{\~n}alosa Esteller}, X. and {Penttil{\"a}}, A. and {Pichon}, B. and {Piersimoni}, A.~M. and {Pineau}, F. -X. and {Plachy}, E. and {Plum}, G. and {Poggio}, E. and {Pr{\v{s}}a}, A. and {Pulone}, L. and {Racero}, E. and {Ragaini}, S. and {Rainer}, M. and {Ramos}, P. and {Ramos-Lerate}, M. and {Re Fiorentin}, P. and {Regibo}, S. and {Richards}, P.~J. and {Rios Diaz}, C. and {Ripepi}, V. and {Riva}, A. and {Rix}, H. -W. and {Rixon}, G. and {Robichon}, N. and {Robin}, A.~C. and {Robin}, C. and {Roelens}, M. and {Rogues}, H.~R.~O. and {Rohrbasser}, L. and {Romero-G{\'o}mez}, M. and {Rowell}, N. and {Royer}, F. and {Ruz Mieres}, D. and {Rybicki}, K.~A. and {Sadowski}, G. and {S{\'a}ez N{\'u}{\~n}ez}, A. and {Sagrist{\`a} Sell{\'e}s}, A. and {Sahlmann}, J. and {Salguero}, E. and {Samaras}, N. and {Sanchez Gimenez}, V. and {Sanna}, N. and {Santove{\~n}a}, R. and {Sarasso}, M. and {Schultheis}, M. and {Sciacca}, E. and {Segol}, M. and {Segovia}, J.~C. and {S{\'e}gransan}, D. and {Semeux}, D. and {Shahaf}, S. and {Siddiqui}, H.~I. and {Siebert}, A. and {Siltala}, L. and {Silvelo}, A. and {Slezak}, E. and {Slezak}, I. and {Smart}, R.~L. and {Snaith}, O.~N. and {Solano}, E. and {Solitro}, F. and {Souami}, D. and {Souchay}, J. and {Spagna}, A. and {Spina}, L. and {Spoto}, F. and {Steele}, I.~A. and {Steidelm{\"u}ller}, H. and {Stephenson}, C.~A. and {S{\"u}veges}, M. and {Surdej}, J. and {Szabados}, L. and {Szegedi-Elek}, E. and {Taris}, F. and {Taylor}, M.~B. and {Tolomei}, L. and {Tonello}, N. and {Torra}, F. and {Torra}, J. and {Torralba Elipe}, G. and {Trabucchi}, M. and {Tsounis}, A.~T. and {Turon}, C. and {Ulla}, A. and {Unger}, N. and {Vaillant}, M.~V. and {van Dillen}, E. and {van Reeven}, W. and {Vanel}, O. and {Vecchiato}, A. and {Viala}, Y. and {Vicente}, D. and {Voutsinas}, S. and {Weiler}, M. and {Wevers}, T. and {Wyrzykowski}, {\L}. and {Yoldas}, A. and {Yvard}, P. and {Zhao}, H. and {Zorec}, J. and {Zucker}, S. and {Zwitter}, T.},
        title = "{Gaia Data Release 3. The extragalactic content}",
      journal = {\aap},
         year = 2023,
        month = jun,
       volume = {674},
          eid = {A41},
        pages = {A41},
          doi = {10.1051/0004-6361/202243232},
archivePrefix = {arXiv},
       eprint = {2206.05681},
 primaryClass = {astro-ph.GA},
       adsurl = {https://ui.adsabs.harvard.edu/abs/2023A&A...674A..41G}
}

@ARTICLE{PanstarrsMain2016,
       author = {{Chambers}, K.~C. and {Magnier}, E.~A. and {Metcalfe}, N. and {Flewelling}, H.~A. and {Huber}, M.~E. and {Waters}, C.~Z. and {Denneau}, L. and {Draper}, P.~W. and {Farrow}, D. and {Finkbeiner}, D.~P. and {Holmberg}, C. and {Koppenhoefer}, J. and {Price}, P.~A. and {Rest}, A. and {Saglia}, R.~P. and {Schlafly}, E.~F. and {Smartt}, S.~J. and {Sweeney}, W. and {Wainscoat}, R.~J. and {Burgett}, W.~S. and {Chastel}, S. and {Grav}, T. and {Heasley}, J.~N. and {Hodapp}, K.~W. and {Jedicke}, R. and {Kaiser}, N. and {Kudritzki}, R. -P. and {Luppino}, G.~A. and {Lupton}, R.~H. and {Monet}, D.~G. and {Morgan}, J.~S. and {Onaka}, P.~M. and {Shiao}, B. and {Stubbs}, C.~W. and {Tonry}, J.~L. and {White}, R. and {Ba{\~n}ados}, E. and {Bell}, E.~F. and {Bender}, R. and {Bernard}, E.~J. and {Boegner}, M. and {Boffi}, F. and {Botticella}, M.~T. and {Calamida}, A. and {Casertano}, S. and {Chen}, W. -P. and {Chen}, X. and {Cole}, S. and {Deacon}, N. and {Frenk}, C. and {Fitzsimmons}, A. and {Gezari}, S. and {Gibbs}, V. and {Goessl}, C. and {Goggia}, T. and {Gourgue}, R. and {Goldman}, B. and {Grant}, P. and {Grebel}, E.~K. and {Hambly}, N.~C. and {Hasinger}, G. and {Heavens}, A.~F. and {Heckman}, T.~M. and {Henderson}, R. and {Henning}, T. and {Holman}, M. and {Hopp}, U. and {Ip}, W. -H. and {Isani}, S. and {Jackson}, M. and {Keyes}, C.~D. and {Koekemoer}, A.~M. and {Kotak}, R. and {Le}, D. and {Liska}, D. and {Long}, K.~S. and {Lucey}, J.~R. and {Liu}, M. and {Martin}, N.~F. and {Masci}, G. and {McLean}, B. and {Mindel}, E. and {Misra}, P. and {Morganson}, E. and {Murphy}, D.~N.~A. and {Obaika}, A. and {Narayan}, G. and {Nieto-Santisteban}, M.~A. and {Norberg}, P. and {Peacock}, J.~A. and {Pier}, E.~A. and {Postman}, M. and {Primak}, N. and {Rae}, C. and {Rai}, A. and {Riess}, A. and {Riffeser}, A. and {Rix}, H.~W. and {R{\"o}ser}, S. and {Russel}, R. and {Rutz}, L. and {Schilbach}, E. and {Schultz}, A.~S.~B. and {Scolnic}, D. and {Strolger}, L. and {Szalay}, A. and {Seitz}, S. and {Small}, E. and {Smith}, K.~W. and {Soderblom}, D.~R. and {Taylor}, P. and {Thomson}, R. and {Taylor}, A.~N. and {Thakar}, A.~R. and {Thiel}, J. and {Thilker}, D. and {Unger}, D. and {Urata}, Y. and {Valenti}, J. and {Wagner}, J. and {Walder}, T. and {Walter}, F. and {Watters}, S.~P. and {Werner}, S. and {Wood-Vasey}, W.~M. and {Wyse}, R.},
        title = "{The Pan-STARRS1 Surveys}",
      journal = {arXiv e-prints},
         year = 2016,
        month = dec,
          eid = {arXiv:1612.05560},
        pages = {arXiv:1612.05560},
          doi = {10.48550/arXiv.1612.05560},
archivePrefix = {arXiv},
       eprint = {1612.05560},
 primaryClass = {astro-ph.IM},
       adsurl = {https://ui.adsabs.harvard.edu/abs/2016arXiv161205560C}
}

@ARTICLE{PanstarrsProcessing2020,
       author = {{Magnier}, Eugene A. and {Chambers}, K.~C. and {Flewelling}, H.~A. and {Hoblitt}, J.~C. and {Huber}, M.~E. and {Price}, P.~A. and {Sweeney}, W.~E. and {Waters}, C.~Z. and {Denneau}, L. and {Draper}, P.~W. and {Hodapp}, K.~W. and {Jedicke}, R. and {Kaiser}, N. and {Kudritzki}, R. -P. and {Metcalfe}, N. and {Stubbs}, C.~W. and {Wainscoat}, R.~J.},
        title = "{The Pan-STARRS Data-processing System}",
      journal = {\apjs},
         year = 2020,
        month = nov,
       volume = {251},
       number = {1},
          eid = {3},
        pages = {3},
          doi = {10.3847/1538-4365/abb829},
archivePrefix = {arXiv},
       eprint = {1612.05240},
 primaryClass = {astro-ph.IM},
       adsurl = {https://ui.adsabs.harvard.edu/abs/2020ApJS..251....3M}
}

@ARTICLE{PanstarrsImageProcessing2020,
       author = {{Waters}, C.~Z. and {Magnier}, E.~A. and {Price}, P.~A. and {Chambers}, K.~C. and {Burgett}, W.~S. and {Draper}, P.~W. and {Flewelling}, H.~A. and {Hodapp}, K.~W. and {Huber}, M.~E. and {Jedicke}, R. and {Kaiser}, N. and {Kudritzki}, R. -P. and {Lupton}, R.~H. and {Metcalfe}, N. and {Rest}, A. and {Sweeney}, W.~E. and {Tonry}, J.~L. and {Wainscoat}, R.~J. and {Wood-Vasey}, W.~M.},
        title = "{Pan-STARRS Pixel Processing: Detrending, Warping, Stacking}",
      journal = {\apjs},
         year = 2020,
        month = nov,
       volume = {251},
       number = {1},
          eid = {4},
        pages = {4},
          doi = {10.3847/1538-4365/abb82b},
archivePrefix = {arXiv},
       eprint = {1612.05245},
 primaryClass = {astro-ph.IM},
       adsurl = {https://ui.adsabs.harvard.edu/abs/2020ApJS..251....4W}
}

@ARTICLE{PanstarrsDatabase2020,
       author = {{Flewelling}, H.~A. and {Magnier}, E.~A. and {Chambers}, K.~C. and {Heasley}, J.~N. and {Holmberg}, C. and {Huber}, M.~E. and {Sweeney}, W. and {Waters}, C.~Z. and {Calamida}, A. and {Casertano}, S. and {Chen}, X. and {Farrow}, D. and {Hasinger}, G. and {Henderson}, R. and {Long}, K.~S. and {Metcalfe}, N. and {Narayan}, G. and {Nieto-Santisteban}, M.~A. and {Norberg}, P. and {Rest}, A. and {Saglia}, R.~P. and {Szalay}, A. and {Thakar}, A.~R. and {Tonry}, J.~L. and {Valenti}, J. and {Werner}, S. and {White}, R. and {Denneau}, L. and {Draper}, P.~W. and {Hodapp}, K.~W. and {Jedicke}, R. and {Kaiser}, N. and {Kudritzki}, R.~P. and {Price}, P.~A. and {Wainscoat}, R.~J. and {Chastel}, S. and {McLean}, B. and {Postman}, M. and {Shiao}, B.},
        title = "{The Pan-STARRS1 Database and Data Products}",
      journal = {\apjs},
         year = 2020,
        month = nov,
       volume = {251},
       number = {1},
          eid = {7},
        pages = {7},
          doi = {10.3847/1538-4365/abb82d},
archivePrefix = {arXiv},
       eprint = {1612.05243},
 primaryClass = {astro-ph.IM},
       adsurl = {https://ui.adsabs.harvard.edu/abs/2020ApJS..251....7F}
}

@ARTICLE{Milliquas7.2,
       author = {{Flesch}, Eric Wim},
        title = "{The Million Quasars (Milliquas) v7.2 Catalogue, now with VLASS associations. The inclusion of SDSS-DR16Q quasars is detailed}",
      journal = {arXiv e-prints},
         year = 2021,
        month = may,
          eid = {arXiv:2105.12985},
        pages = {arXiv:2105.12985},
          doi = {10.48550/arXiv.2105.12985},
archivePrefix = {arXiv},
       eprint = {2105.12985},
 primaryClass = {astro-ph.GA},
       adsurl = {https://ui.adsabs.harvard.edu/abs/2021arXiv210512985F}
}

@ARTICLE{Milliquas8,
       author = {{Flesch}, Eric Wim},
        title = "{The Million Quasars (Milliquas) Catalogue, v8}",
      journal = {The Open Journal of Astrophysics},
         year = 2023,
        month = dec,
       volume = {6},
          eid = {49},
        pages = {49},
          doi = {10.21105/astro.2308.01505},
archivePrefix = {arXiv},
       eprint = {2308.01505},
 primaryClass = {astro-ph.GA},
       adsurl = {https://ui.adsabs.harvard.edu/abs/2023OJAp....6E..49F}
}

@ARTICLE{Milliquas6.4,
       author = {{Flesch}, Eric W.},
        title = "{The Million Quasars (Milliquas) Catalogue, v6.4}",
      journal = {arXiv e-prints},
         year = 2019,
        month = dec,
          eid = {arXiv:1912.05614},
        pages = {arXiv:1912.05614},
          doi = {10.48550/arXiv.1912.05614},
archivePrefix = {arXiv},
       eprint = {1912.05614},
 primaryClass = {astro-ph.GA},
       adsurl = {https://ui.adsabs.harvard.edu/abs/2019arXiv191205614F}
}

@ARTICLE{DESI,
       author = {{Dey}, Arjun and {Schlegel}, David J. and {Lang}, Dustin and {Blum}, Robert and {Burleigh}, Kaylan and {Fan}, Xiaohui and {Findlay}, Joseph R. and {Finkbeiner}, Doug and {Herrera}, David and {Juneau}, St{\'e}phanie and {Landriau}, Martin and {Levi}, Michael and {McGreer}, Ian and {Meisner}, Aaron and {Myers}, Adam D. and {Moustakas}, John and {Nugent}, Peter and {Patej}, Anna and {Schlafly}, Edward F. and {Walker}, Alistair R. and {Valdes}, Francisco and {Weaver}, Benjamin A. and {Y{\`e}che}, Christophe and {Zou}, Hu and {Zhou}, Xu and {Abareshi}, Behzad and {Abbott}, T.~M.~C. and {Abolfathi}, Bela and {Aguilera}, C. and {Alam}, Shadab and {Allen}, Lori and {Alvarez}, A. and {Annis}, James and {Ansarinejad}, Behzad and {Aubert}, Marie and {Beechert}, Jacqueline and {Bell}, Eric F. and {BenZvi}, Segev Y. and {Beutler}, Florian and {Bielby}, Richard M. and {Bolton}, Adam S. and {Brice{\~n}o}, C{\'e}sar and {Buckley-Geer}, Elizabeth J. and {Butler}, Karen and {Calamida}, Annalisa and {Carlberg}, Raymond G. and {Carter}, Paul and {Casas}, Ricard and {Castander}, Francisco J. and {Choi}, Yumi and {Comparat}, Johan and {Cukanovaite}, Elena and {Delubac}, Timoth{\'e}e and {DeVries}, Kaitlin and {Dey}, Sharmila and {Dhungana}, Govinda and {Dickinson}, Mark and {Ding}, Zhejie and {Donaldson}, John B. and {Duan}, Yutong and {Duckworth}, Christopher J. and {Eftekharzadeh}, Sarah and {Eisenstein}, Daniel J. and {Etourneau}, Thomas and {Fagrelius}, Parker A. and {Farihi}, Jay and {Fitzpatrick}, Mike and {Font-Ribera}, Andreu and {Fulmer}, Leah and {G{\"a}nsicke}, Boris T. and {Gaztanaga}, Enrique and {George}, Koshy and {Gerdes}, David W. and {Gontcho}, Satya Gontcho A. and {Gorgoni}, Claudio and {Green}, Gregory and {Guy}, Julien and {Harmer}, Diane and {Hernandez}, M. and {Honscheid}, Klaus and {Huang}, Lijuan Wendy and {James}, David J. and {Jannuzi}, Buell T. and {Jiang}, Linhua and {Joyce}, Richard and {Karcher}, Armin and {Karkar}, Sonia and {Kehoe}, Robert and {Kneib}, Jean-Paul and {Kueter-Young}, Andrea and {Lan}, Ting-Wen and {Lauer}, Tod R. and {Le Guillou}, Laurent and {Le Van Suu}, Auguste and {Lee}, Jae Hyeon and {Lesser}, Michael and {Perreault Levasseur}, Laurence and {Li}, Ting S. and {Mann}, Justin L. and {Marshall}, Robert and {Mart{\'\i}nez-V{\'a}zquez}, C.~E. and {Martini}, Paul and {du Mas des Bourboux}, H{\'e}lion and {McManus}, Sean and {Meier}, Tobias Gabriel and {M{\'e}nard}, Brice and {Metcalfe}, Nigel and {Mu{\~n}oz-Guti{\'e}rrez}, Andrea and {Najita}, Joan and {Napier}, Kevin and {Narayan}, Gautham and {Newman}, Jeffrey A. and {Nie}, Jundan and {Nord}, Brian and {Norman}, Dara J. and {Olsen}, Knut A.~G. and {Paat}, Anthony and {Palanque-Delabrouille}, Nathalie and {Peng}, Xiyan and {Poppett}, Claire L. and {Poremba}, Megan R. and {Prakash}, Abhishek and {Rabinowitz}, David and {Raichoor}, Anand and {Rezaie}, Mehdi and {Robertson}, A.~N. and {Roe}, Natalie A. and {Ross}, Ashley J. and {Ross}, Nicholas P. and {Rudnick}, Gregory and {Safonova}, Sasha and {Saha}, Abhijit and {S{\'a}nchez}, F. Javier and {Savary}, Elodie and {Schweiker}, Heidi and {Scott}, Adam and {Seo}, Hee-Jong and {Shan}, Huanyuan and {Silva}, David R. and {Slepian}, Zachary and {Soto}, Christian and {Sprayberry}, David and {Staten}, Ryan and {Stillman}, Coley M. and {Stupak}, Robert J. and {Summers}, David L. and {Sien Tie}, Suk and {Tirado}, H. and {Vargas-Maga{\~n}a}, Mariana and {Vivas}, A. Katherina and {Wechsler}, Risa H. and {Williams}, Doug and {Yang}, Jinyi and {Yang}, Qian and {Yapici}, Tolga and {Zaritsky}, Dennis and {Zenteno}, A. and {Zhang}, Kai and {Zhang}, Tianmeng and {Zhou}, Rongpu and {Zhou}, Zhimin},
        title = "{Overview of the DESI Legacy Imaging Surveys}",
      journal = {\aj},
         year = 2019,
        month = may,
       volume = {157},
       number = {5},
          eid = {168},
        pages = {168},
          doi = {10.3847/1538-3881/ab089d},
archivePrefix = {arXiv},
       eprint = {1804.08657},
 primaryClass = {astro-ph.IM},
       adsurl = {https://ui.adsabs.harvard.edu/abs/2019AJ....157..168D}
}

@ARTICLE{VAT,
       author = {{Miyato}, Takeru and {Maeda}, Shin-ichi and {Koyama}, Masanori and {Ishii}, Shin},
        title = "{Virtual Adversarial Training: A Regularization Method for Supervised and Semi-Supervised Learning}",
      journal = {arXiv e-prints},
         year = 2017,
        month = apr,
          eid = {arXiv:1704.03976},
        pages = {arXiv:1704.03976},
          doi = {10.48550/arXiv.1704.03976},
archivePrefix = {arXiv},
       eprint = {1704.03976},
 primaryClass = {stat.ML},
       adsurl = {https://ui.adsabs.harvard.edu/abs/2017arXiv170403976M}
}

@article{Autoencoder,
author = {G. E. Hinton  and R. R. Salakhutdinov },
title = {Reducing the Dimensionality of Data with Neural Networks},
journal = {Science},
volume = {313},
number = {5786},
pages = {504-507},
year = {2006},
doi = {10.1126/science.1127647},
URL = {https://www.science.org/doi/abs/10.1126/science.1127647},
eprint = {https://www.science.org/doi/pdf/10.1126/science.1127647}}

@ARTICLE{RandomForest,
       author = {{Breiman}, Leo},
        title = "{Random Forests.}",
      journal = {Machine Learning},
         year = 2001,
        month = jan,
       volume = {45},
        pages = {5-32},
          doi = {10.1023/A:1010933404324},
       adsurl = {https://ui.adsabs.harvard.edu/abs/2001MachL..45....5B}
}

@inproceedings{LeCun1989,
 author = {LeCun, Yann and Boser, Bernhard and Denker, John and Henderson, Donnie and Howard, R. and Hubbard, Wayne and Jackel, Lawrence},
 booktitle = {Advances in Neural Information Processing Systems},
 editor = {D. Touretzky},
 pages = {},
 publisher = {Morgan-Kaufmann},
 title = {Handwritten Digit Recognition with a Back-Propagation Network},
 url = {https://proceedings.neurips.cc/paper_files/paper/1989/file/53c3bce66e43be4f209556518c2fcb54-Paper.pdf},
 volume = {2},
 year = {1989},
 series = {}
}

@INPROCEEDINGS{Zeiler2010,
  author={Zeiler, Matthew D. and Krishnan, Dilip and Taylor, Graham W. and Fergus, Rob},
  booktitle={2010 IEEE Computer Society Conference on Computer Vision and Pattern Recognition}, 
  title={Deconvolutional networks}, 
  year={2010},
  volume={},
  number={},
  pages={2528-2535},
  doi={10.1109/CVPR.2010.5539957}}

@ARTICLE{Kingma2013,
       author = {{Kingma}, Diederik P and {Welling}, Max},
        title = "{Auto-Encoding Variational Bayes}",
      journal = {arXiv e-prints},
         year = 2013,
        month = dec,
          eid = {arXiv:1312.6114},
        pages = {arXiv:1312.6114},
          doi = {10.48550/arXiv.1312.6114},
archivePrefix = {arXiv},
       eprint = {1312.6114},
 primaryClass = {stat.ML},
       adsurl = {https://ui.adsabs.harvard.edu/abs/2013arXiv1312.6114K}
}

@article{Kullback1951,
author = {S. Kullback and R. A. Leibler},
title = {{On Information and Sufficiency}},
volume = {22},
journal = {The Annals of Mathematical Statistics},
number = {1},
publisher = {Institute of Mathematical Statistics},
pages = {79 -- 86},
year = {1951},
doi = {10.1214/aoms/1177729694},
URL = {https://doi.org/10.1214/aoms/1177729694}
}

@inproceedings{Higgins2017,
title={beta-{VAE}: Learning Basic Visual Concepts with a Constrained Variational Framework},
author={Irina Higgins and Loic Matthey and Arka Pal and Christopher Burgess and Xavier Glorot and Matthew Botvinick and Shakir Mohamed and Alexander Lerchner},
booktitle={International Conference on Learning Representations},
year={2017},
url={https://openreview.net/forum?id=Sy2fzU9gl}
}

@article{IsolationForest,
author = {Liu, Fei Tony and Ting, Kai Ming and Zhou, Zhi-Hua},
title = {Isolation-Based Anomaly Detection},
year = {2012},
issue_date = {March 2012},
publisher = {Association for Computing Machinery},
address = {New York, NY, USA},
volume = {6},
number = {1},
issn = {1556-4681},
url = {https://doi.org/10.1145/2133360.2133363},
doi = {10.1145/2133360.2133363},
journal = {ACM Trans. Knowl. Discov. Data},
month = {mar},
articleno = {3},
numpages = {39}
}

@article{LogisticRegression,
    author = {Joseph Berkson},
    title = {Application of the Logistic Function to Bio-Assay},
    journal = {Journal of the American Statistical Association},
    volume = {39},
    number = {227},
    pages = {357--365},
    year = {1944},
    publisher = {Taylor \& Francis},
    doi = {10.1080/01621459.1944.10500699},
    URL = {https://doi.org/10.1080/01621459.1944.10500699},
    eprint = {https://doi.org/10.1080/01621459.1944.10500699}
}

@article{GradientBoosting,
author = {Jerome H. Friedman},
title = {{Greedy function approximation: A gradient boosting machine.}},
volume = {29},
journal = {The Annals of Statistics},
number = {5},
publisher = {Institute of Mathematical Statistics},
pages = {1189 -- 1232},
year = {2001},
doi = {10.1214/aos/1013203451},
URL = {https://doi.org/10.1214/aos/1013203451}
}

@book{SVM,
	author = {Vladimir Vapnik},
	editor = {},
	publisher = {Springer: New York},
	title = {The Nature of Statistical Learning Theory},
	year = {1995}
}

@article{K-SVM,
author = {Cortes, Corinna and Vapnik, Vladimir},
title = {Support-Vector Networks},
year = {1995},
issue_date = {Sept. 1995},
publisher = {Kluwer Academic Publishers},
address = {USA},
volume = {20},
number = {3},
issn = {0885-6125},
url = {https://doi.org/10.1023/A:1022627411411},
doi = {10.1023/A:1022627411411},
journal = {Mach. Learn.},
month = {sep},
pages = {273–297},
numpages = {25}
}

@BOOK{GP,
       author = {{Rasmussen}, Carl Edward and {Williams}, Christopher K.~I.},
        title = "{Gaussian Processes for Machine Learning}",
         year = 2006,
       adsurl = {https://ui.adsabs.harvard.edu/abs/2006gpml.book.....R},
    publisher = "{The MIT Press}",
         ISBN = "{0-262-18253-X}"
}

@ARTICLE{Ioffe2015,
       author = {{Ioffe}, Sergey and {Szegedy}, Christian},
        title = "{Batch Normalization: Accelerating Deep Network Training by Reducing Internal Covariate Shift}",
      journal = {arXiv e-prints},
         year = 2015,
        month = feb,
          eid = {arXiv:1502.03167},
        pages = {arXiv:1502.03167},
          doi = {10.48550/arXiv.1502.03167},
archivePrefix = {arXiv},
       eprint = {1502.03167},
 primaryClass = {cs.LG},
       adsurl = {https://ui.adsabs.harvard.edu/abs/2015arXiv150203167I}
}

@ARTICLE{Akhazhanov2022,
       author = {{Akhazhanov}, A. and {More}, A. and {Amini}, A. and {Hazlett}, C. and {Treu}, T. and {Birrer}, S. and {Shajib}, A. and {Liao}, K. and {Lemon}, C. and {Agnello}, A. and {Nord}, B. and {Aguena}, M. and {Allam}, S. and {Andrade-Oliveira}, F. and {Annis}, J. and {Brooks}, D. and {Buckley-Geer}, E. and {Burke}, D.~L. and {Carnero Rosell}, A. and {Carrasco Kind}, M. and {Carretero}, J. and {Choi}, A. and {Conselice}, C. and {Costanzi}, M. and {da Costa}, L.~N. and {Pereira}, M.~E.~S. and {De Vicente}, J. and {Desai}, S. and {Dietrich}, J.~P. and {Doel}, P. and {Everett}, S. and {Ferrero}, I. and {Finley}, D.~A. and {Flaugher}, B. and {Frieman}, J. and {Garc{\'\i}a-Bellido}, J. and {Gerdes}, D.~W. and {Gruen}, D. and {Gruendl}, R.~A. and {Gschwend}, J. and {Gutierrez}, G. and {Hinton}, S.~R. and {Hollowood}, D.~L. and {Honscheid}, K. and {James}, D.~J. and {Kim}, A.~G. and {Kuehn}, K. and {Kuropatkin}, N. and {Lahav}, O. and {Lima}, M. and {Lin}, H. and {Maia}, M.~A.~G. and {March}, M. and {Menanteau}, F. and {Miquel}, R. and {Morgan}, R. and {Palmese}, A. and {Paz-Chinch{\'o}n}, F. and {Pieres}, A. and {Plazas Malag{\'o}n}, A.~A. and {Sanchez}, E. and {Scarpine}, V. and {Serrano}, S. and {Sevilla-Noarbe}, I. and {Smith}, M. and {Soares-Santos}, M. and {Suchyta}, E. and {Swanson}, M.~E.~C. and {Tarle}, G. and {To}, C. and {Varga}, T.~N. and {Weller}, J. and {(DES Collaboration)}},
        title = "{Finding quadruply imaged quasars with machine learning - I. Methods}",
      journal = {\mnras},
         year = 2022,
        month = jun,
       volume = {513},
       number = {2},
        pages = {2407-2421},
          doi = {10.1093/mnras/stac925},
archivePrefix = {arXiv},
       eprint = {2109.09781},
 primaryClass = {astro-ph.CO},
       adsurl = {https://ui.adsabs.harvard.edu/abs/2022MNRAS.513.2407A}
}

@ARTICLE{Einstein1916,
       author = {{Einstein}, A.},
        title = "{Die Grundlage der allgemeinen Relativit{\"a}tstheorie}",
      journal = {Annalen der Physik},
         year = 1916,
        month = jan,
       volume = {354},
       number = {7},
        pages = {769-822},
          doi = {10.1002/andp.19163540702},
       adsurl = {https://ui.adsabs.harvard.edu/abs/1916AnP...354..769E}
}

@ARTICLE{Walsh1979,
       author = {{Walsh}, D. and {Carswell}, R.~F. and {Weymann}, R.~J.},
        title = "{0957+561 A, B: twin quasistellar objects or gravitational lens?}",
      journal = {\nat},
         year = 1979,
        month = may,
       volume = {279},
        pages = {381-384},
          doi = {10.1038/279381a0},
       adsurl = {https://ui.adsabs.harvard.edu/abs/1979Natur.279..381W}
}

@ARTICLE{Rauch2001,
       author = {{Rauch}, Michael and {Sargent}, Wallace L.~W. and {Barlow}, Thomas A.},
        title = "{Small-Scale Structure at High Redshift. II. Physical Properties of the C IV Absorbing Clouds}",
      journal = {\apj},
         year = 2001,
        month = jun,
       volume = {554},
       number = {2},
        pages = {823-840},
          doi = {10.1086/321402},
archivePrefix = {arXiv},
       eprint = {astro-ph/0104216},
 primaryClass = {astro-ph},
       adsurl = {https://ui.adsabs.harvard.edu/abs/2001ApJ...554..823R}
}

@ARTICLE{Rauch2011,
       author = {{Rauch}, Michael and {Haehnelt}, Martin G.},
        title = "{Faint resonantly scattered Ly{\ensuremath{\alpha}} emission from the absorption troughs of damped Ly{\ensuremath{\alpha}} systems at z{\ensuremath{\sim}} 3}",
      journal = {\mnras},
         year = 2011,
        month = mar,
       volume = {412},
       number = {1},
        pages = {L55-L57},
          doi = {10.1111/j.1745-3933.2010.01004.x},
archivePrefix = {arXiv},
       eprint = {1011.4061},
 primaryClass = {astro-ph.CO},
       adsurl = {https://ui.adsabs.harvard.edu/abs/2011MNRAS.412L..55R}
}

@ARTICLE{Smette1995,
       author = {{Smette}, A. and {Robertson}, J.~G. and {Shaver}, P.~A. and {Reimers}, D. and {Wisotzki}, L. and {Koehler}, T.},
        title = "{The gravitational lens candidate HE 1104-1805 and the size of absorption systems.}",
      journal = {\aaps},
         year = 1995,
        month = oct,
       volume = {113},
        pages = {199},
       adsurl = {https://ui.adsabs.harvard.edu/abs/1995A&AS..113..199S}
}

@ARTICLE{Foltz1984,
       author = {{Foltz}, C.~B. and {Weymann}, R.~J. and {Roser}, H. -J. and {Chaffee}, F.~H., Jr.},
        title = "{Improved lower limits on Lyman-alpha forest cloud dimensions and additional evidence supporting the gravitational lens nature of 2345+007 A, B.}",
      journal = {\apjl},
         year = 1984,
        month = jun,
       volume = {281},
        pages = {L1-L4},
          doi = {10.1086/184271},
       adsurl = {https://ui.adsabs.harvard.edu/abs/1984ApJ...281L...1F}
}

@ARTICLE{Meyer2023,
       author = {{Meyer}, Antoine D. and {van Dyk}, David A. and {Tak}, Hyungsuk and {Siemiginowska}, Aneta},
        title = "{TD-CARMA: Painless, Accurate, and Scalable Estimates of Gravitational Lens Time Delays with Flexible CARMA Processes}",
      journal = {\apj},
         year = 2023,
        month = jun,
       volume = {950},
       number = {1},
          eid = {37},
        pages = {37},
          doi = {10.3847/1538-4357/acbea1},
archivePrefix = {arXiv},
       eprint = {2207.09327},
 primaryClass = {astro-ph.IM},
       adsurl = {https://ui.adsabs.harvard.edu/abs/2023ApJ...950...37M}
}

@ARTICLE{Refsdal1964,
       author = {{Refsdal}, S.},
        title = "{On the possibility of determining Hubble's parameter and the masses of galaxies from the gravitational lens effect}",
      journal = {\mnras},
         year = 1964,
        month = jan,
       volume = {128},
        pages = {307},
          doi = {10.1093/mnras/128.4.307},
       adsurl = {https://ui.adsabs.harvard.edu/abs/1964MNRAS.128..307R}
}

@ARTICLE{Weinberger2018,
       author = {{Weinberger}, Rainer and {Springel}, Volker and {Pakmor}, R{\"u}diger and {Nelson}, Dylan and {Genel}, Shy and {Pillepich}, Annalisa and {Vogelsberger}, Mark and {Marinacci}, Federico and {Naiman}, Jill and {Torrey}, Paul and {Hernquist}, Lars},
        title = "{Supermassive black holes and their feedback effects in the IllustrisTNG simulation}",
      journal = {\mnras},
         year = 2018,
        month = sep,
       volume = {479},
       number = {3},
        pages = {4056-4072},
          doi = {10.1093/mnras/sty1733},
archivePrefix = {arXiv},
       eprint = {1710.04659},
 primaryClass = {astro-ph.GA},
       adsurl = {https://ui.adsabs.harvard.edu/abs/2018MNRAS.479.4056W}
}

@ARTICLE{Volonteri2010,
       author = {{Volonteri}, Marta},
        title = "{Formation of supermassive black holes}",
      journal = {\aapr},
         year = 2010,
        month = jul,
       volume = {18},
       number = {3},
        pages = {279-315},
          doi = {10.1007/s00159-010-0029-x},
archivePrefix = {arXiv},
       eprint = {1003.4404},
 primaryClass = {astro-ph.CO},
       adsurl = {https://ui.adsabs.harvard.edu/abs/2010A&ARv..18..279V}
}

@ARTICLE{Krone-Martins2019,
       author = {{Krone-Martins}, A. and {Graham}, M.~J. and {Stern}, D. and {Djorgovski}, S.~G. and {Delchambre}, L. and {Ducourant}, C. and {Teixeira}, R. and {Drake}, A.~J. and {Scarano}, S., Jr. and {Surdej}, J. and {Galluccio}, L. and {Jalan}, P. and {Wertz}, O. and {Kl{\"u}ter}, J. and {Mignard}, F. and {Spindola-Duarte}, C. and {Dobie}, D. and {Slezak}, E. and {Sluse}, D. and {Murphy}, T. and {Boehm}, C. and {Nierenberg}, A.~M. and {Bastian}, U. and {Wambsganss}, J. and {LeCampion}, J. -F.},
        title = "{Gaia GraL: Gaia DR2 Gravitational Lens Systems. V. Doubly-imaged QSOs discovered from entropy and wavelets}",
      journal = {arXiv e-prints},
         year = 2019,
        month = dec,
          eid = {arXiv:1912.08977},
        pages = {arXiv:1912.08977},
          doi = {10.48550/arXiv.1912.08977},
archivePrefix = {arXiv},
       eprint = {1912.08977},
 primaryClass = {astro-ph.GA},
       adsurl = {https://ui.adsabs.harvard.edu/abs/2019arXiv191208977K}
}

@ARTICLE{Ducourant2018,
       author = {{Ducourant}, C. and {Wertz}, O. and {Krone-Martins}, A. and {Teixeira}, R. and {Le Campion}, J. -F. and {Galluccio}, L. and {Kl{\"u}ter}, J. and {Delchambre}, L. and {Surdej}, J. and {Mignard}, F. and {Wambsganss}, J. and {Bastian}, U. and {Graham}, M.~J. and {Djorgovski}, S.~G. and {Slezak}, E.},
        title = "{Gaia GraL: Gaia DR2 gravitational lens systems. II. The known multiply imaged quasars}",
      journal = {\aap},
         year = 2018,
        month = oct,
       volume = {618},
          eid = {A56},
        pages = {A56},
          doi = {10.1051/0004-6361/201833480},
       adsurl = {https://ui.adsabs.harvard.edu/abs/2018A&A...618A..56D}
}

@ARTICLE{Lemon2019,
       author = {{Lemon}, Cameron A. and {Auger}, Matthew W. and {McMahon}, Richard G.},
        title = "{Gravitationally lensed quasars in Gaia - III. 22 new lensed quasars from Gaia data release 2}",
      journal = {\mnras},
         year = 2019,
        month = mar,
       volume = {483},
       number = {3},
        pages = {4242-4258},
          doi = {10.1093/mnras/sty3366},
archivePrefix = {arXiv},
       eprint = {1810.04480},
 primaryClass = {astro-ph.GA},
       adsurl = {https://ui.adsabs.harvard.edu/abs/2019MNRAS.483.4242L}
}

@ARTICLE{Einstein1936,
       author = {{Einstein}, Albert},
        title = "{Lens-Like Action of a Star by the Deviation of Light in the Gravitational Field}",
      journal = {Science},
         year = 1936,
        month = dec,
       volume = {84},
       number = {2188},
        pages = {506-507},
          doi = {10.1126/science.84.2188.506},
       adsurl = {https://ui.adsabs.harvard.edu/abs/1936Sci....84..506E}
}

@ARTICLE{Zwicky1937,
       author = {{Zwicky}, F.},
        title = "{Nebulae as Gravitational Lenses}",
      journal = {Physical Review},
         year = 1937,
        month = feb,
       volume = {51},
       number = {4},
        pages = {290-290},
          doi = {10.1103/PhysRev.51.290},
       adsurl = {https://ui.adsabs.harvard.edu/abs/1937PhRv...51..290Z}
}

@ARTICLE{Sanchez2021a,
       author = {{S{\'a}nchez-S{\'a}ez}, P. and {Reyes}, I. and {Valenzuela}, C. and {F{\"o}rster}, F. and {Eyheramendy}, S. and {Elorrieta}, F. and {Bauer}, F.~E. and {Cabrera-Vives}, G. and {Est{\'e}vez}, P.~A. and {Catelan}, M. and {Pignata}, G. and {Huijse}, P. and {De Cicco}, D. and {Ar{\'e}valo}, P. and {Carrasco-Davis}, R. and {Abril}, J. and {Kurtev}, R. and {Borissova}, J. and {Arredondo}, J. and {Castillo-Navarrete}, E. and {Rodriguez}, D. and {Ruz-Mieres}, D. and {Moya}, A. and {Sabatini-Gacit{\'u}a}, L. and {Sep{\'u}lveda-Cobo}, C. and {Camacho-I{\~n}iguez}, E.},
        title = "{Alert Classification for the ALeRCE Broker System: The Light Curve Classifier}",
      journal = {\aj},
         year = 2021,
        month = mar,
       volume = {161},
       number = {3},
          eid = {141},
        pages = {141},
          doi = {10.3847/1538-3881/abd5c1},
archivePrefix = {arXiv},
       eprint = {2008.03311},
 primaryClass = {astro-ph.IM},
       adsurl = {https://ui.adsabs.harvard.edu/abs/2021AJ....161..141S}
}

@ARTICLE{Carrasco2021,
       author = {{Carrasco-Davis}, R. and {Reyes}, E. and {Valenzuela}, C. and {F{\"o}rster}, F. and {Est{\'e}vez}, P.~A. and {Pignata}, G. and {Bauer}, F.~E. and {Reyes}, I. and {S{\'a}nchez-S{\'a}ez}, P. and {Cabrera-Vives}, G. and {Eyheramendy}, S. and {Catelan}, M. and {Arredondo}, J. and {Castillo-Navarrete}, E. and {Rodr{\'\i}guez-Mancini}, D. and {Ruz-Mieres}, D. and {Moya}, A. and {Sabatini-Gacit{\'u}a}, L. and {Sep{\'u}lveda-Cobo}, C. and {Mahabal}, A.~A. and {Silva-Farf{\'a}n}, J. and {Camacho-I{\~n}iguez}, E. and {Galbany}, L.},
        title = "{Alert Classification for the ALeRCE Broker System: The Real-time Stamp Classifier}",
      journal = {\aj},
         year = 2021,
        month = dec,
       volume = {162},
       number = {6},
          eid = {231},
        pages = {231},
          doi = {10.3847/1538-3881/ac0ef1},
archivePrefix = {arXiv},
       eprint = {2008.03309},
 primaryClass = {astro-ph.IM},
       adsurl = {https://ui.adsabs.harvard.edu/abs/2021AJ....162..231C}
}

@ARTICLE{Perez2023,
       author = {{Perez-Carrasco}, Manuel and {Cabrera-Vives}, Guillermo and {Hernandez-Garc{\'\i}a}, Lorena and {F{\"o}rster}, F. and {Sanchez-Saez}, Paula and {Mu{\~n}oz Arancibia}, Alejandra M. and {Arredondo}, Javier and {Astorga}, Nicol{\'a}s and {Bauer}, Franz E. and {Bayo}, Amelia and {Catelan}, M. and {Dastidar}, Raya and {Est{\'e}vez}, P.~A. and {Lira}, Paulina and {Pignata}, Giuliano},
        title = "{Alert Classification for the ALeRCE Broker System: The Anomaly Detector}",
      journal = {\aj},
         year = 2023,
        month = oct,
       volume = {166},
       number = {4},
          eid = {151},
        pages = {151},
          doi = {10.3847/1538-3881/ace0c1},
       adsurl = {https://ui.adsabs.harvard.edu/abs/2023AJ....166..151P}
}

@ARTICLE{Alerce2021,
       author = {{F{\"o}rster}, F. and {Cabrera-Vives}, G. and {Castillo-Navarrete}, E. and {Est{\'e}vez}, P.~A. and {S{\'a}nchez-S{\'a}ez}, P. and {Arredondo}, J. and {Bauer}, F.~E. and {Carrasco-Davis}, R. and {Catelan}, M. and {Elorrieta}, F. and {Eyheramendy}, S. and {Huijse}, P. and {Pignata}, G. and {Reyes}, E. and {Reyes}, I. and {Rodr{\'\i}guez-Mancini}, D. and {Ruz-Mieres}, D. and {Valenzuela}, C. and {{\'A}lvarez-Maldonado}, I. and {Astorga}, N. and {Borissova}, J. and {Clocchiatti}, A. and {De Cicco}, D. and {Donoso-Oliva}, C. and {Hern{\'a}ndez-Garc{\'\i}a}, L. and {Graham}, M.~J. and {Jord{\'a}n}, A. and {Kurtev}, R. and {Mahabal}, A. and {Maureira}, J.~C. and {Mu{\~n}oz-Arancibia}, A. and {Molina-Ferreiro}, R. and {Moya}, A. and {Palma}, W. and {P{\'e}rez-Carrasco}, M. and {Protopapas}, P. and {Romero}, M. and {Sabatini-Gacitua}, L. and {S{\'a}nchez}, A. and {San Mart{\'\i}n}, J. and {Sep{\'u}lveda-Cobo}, C. and {Vera}, E. and {Vergara}, J.~R.},
        title = "{The Automatic Learning for the Rapid Classification of Events (ALeRCE) Alert Broker}",
      journal = {\aj},
         year = 2021,
        month = may,
       volume = {161},
       number = {5},
          eid = {242},
        pages = {242},
          doi = {10.3847/1538-3881/abe9bc},
archivePrefix = {arXiv},
       eprint = {2008.03303},
 primaryClass = {astro-ph.IM},
       adsurl = {https://ui.adsabs.harvard.edu/abs/2021AJ....161..242F}
}

@ARTICLE{Sanchez2021b,
       author = {{S{\'a}nchez-S{\'a}ez}, P. and {Lira}, H. and {Mart{\'\i}}, L. and {S{\'a}nchez-Pi}, N. and {Arredondo}, J. and {Bauer}, F.~E. and {Bayo}, A. and {Cabrera-Vives}, G. and {Donoso-Oliva}, C. and {Est{\'e}vez}, P.~A. and {Eyheramendy}, S. and {F{\"o}rster}, F. and {Hern{\'a}ndez-Garc{\'\i}a}, L. and {Arancibia}, A.~M. Mu{\~n}oz and {P{\'e}rez-Carrasco}, M. and {Sep{\'u}lveda}, M. and {Vergara}, J.~R.},
        title = "{Searching for Changing-state AGNs in Massive Data Sets. I. Applying Deep Learning and Anomaly-detection Techniques to Find AGNs with Anomalous Variability Behaviors}",
      journal = {\aj},
         year = 2021,
        month = nov,
       volume = {162},
       number = {5},
          eid = {206},
        pages = {206},
          doi = {10.3847/1538-3881/ac1426},
archivePrefix = {arXiv},
       eprint = {2106.07660},
 primaryClass = {astro-ph.IM},
       adsurl = {https://ui.adsabs.harvard.edu/abs/2021AJ....162..206S}
}

@ARTICLE{Forster2022,
       author = {{F{\"o}rster}, Francisco and {Mu{\~n}oz Arancibia}, Alejandra M. and {Reyes-Jainaga}, Ignacio and {Gagliano}, Alexander and {Britt}, Dylan and {Cuellar-Carrillo}, Sara and {Figueroa-Tapia}, Felipe and {Polzin}, Ava and {Yousef}, Yara and {Arredondo}, Javier and {Rodr{\'\i}guez-Mancini}, Diego and {Correa-Orellana}, Javier and {Bayo}, Amelia and {Bauer}, Franz E. and {Catelan}, M{\'a}rcio and {Cabrera-Vives}, Guillermo and {Dastidar}, Raya and {Est{\'e}vez}, Pablo A. and {Pignata}, Giuliano and {Hern{\'a}ndez-Garc{\'\i}a}, Lorena and {Huijse}, Pablo and {Reyes}, Esteban and {S{\'a}nchez-S{\'a}ez}, Paula and {Ram{\'\i}rez}, Mauricio and {Grand{\'o}n}, Daniela and {Pineda-Garc{\'\i}a}, Jonathan and {Chabour-Barra}, Francisca and {Silva-Farf{\'a}n}, Javier},
        title = "{DELIGHT: Deep Learning Identification of Galaxy Hosts of Transients using Multiresolution Images}",
      journal = {\aj},
         year = 2022,
        month = nov,
       volume = {164},
       number = {5},
          eid = {195},
        pages = {195},
          doi = {10.3847/1538-3881/ac912a},
archivePrefix = {arXiv},
       eprint = {2208.04310},
 primaryClass = {astro-ph.IM},
       adsurl = {https://ui.adsabs.harvard.edu/abs/2022AJ....164..195F}
}

@ARTICLE{PyTorch,
       author = {{Paszke}, Adam and {Gross}, Sam and {Massa}, Francisco and {Lerer}, Adam and {Bradbury}, James and {Chanan}, Gregory and {Killeen}, Trevor and {Lin}, Zeming and {Gimelshein}, Natalia and {Antiga}, Luca and {Desmaison}, Alban and {K{\"o}pf}, Andreas and {Yang}, Edward and {DeVito}, Zach and {Raison}, Martin and {Tejani}, Alykhan and {Chilamkurthy}, Sasank and {Steiner}, Benoit and {Fang}, Lu and {Bai}, Junjie and {Chintala}, Soumith},
        title = "{PyTorch: An Imperative Style, High-Performance Deep Learning Library}",
      journal = {arXiv e-prints},
         year = 2019,
        month = dec,
          eid = {arXiv:1912.01703},
        pages = {arXiv:1912.01703},
          doi = {10.48550/arXiv.1912.01703},
archivePrefix = {arXiv},
       eprint = {1912.01703},
 primaryClass = {cs.LG},
       adsurl = {https://ui.adsabs.harvard.edu/abs/2019arXiv191201703P}
}

@article{scikit-learn,
  title={Scikit-learn: Machine Learning in {P}ython},
  author={Pedregosa, F. and Varoquaux, G. and Gramfort, A. and Michel, V.
          and Thirion, B. and Grisel, O. and Blondel, M. and Prettenhofer, P.
          and Weiss, R. and Dubourg, V. and Vanderplas, J. and Passos, A. and
          Cournapeau, D. and Brucher, M. and Perrot, M. and Duchesnay, E.},
  journal={Journal of Machine Learning Research},
  volume={12},
  pages={2825--2830},
  year={2011}
}

@article{PIL,
  title={Image Processing in Python},
  author={Umesh, P},
  journal={CSI Communications},
  volume={23},
  year={2012},
  publisher={Citeseer}
}

@article{scikit-image,
 title = {scikit-image: image processing in {P}ython},
 author = {van der Walt, {S}t\'efan and {S}ch\"onberger, {J}ohannes {L}. and
           {Nunez-Iglesias}, {J}uan and {B}oulogne, {F}ran\c{c}ois and {W}arner,
           {J}oshua {D}. and {Y}ager, {N}eil and {G}ouillart, {E}mmanuelle and
           {Y}u, {T}ony and the scikit-image contributors},
 year = {2014},
 month = {6},
 volume = {2},
 pages = {e453},
 journal = {PeerJ},
 issn = {2167-8359},
 url = {https://doi.org/10.7717/peerj.453},
 doi = {10.7717/peerj.453}
}

@ARTICLE{NumPy,
       author = {{Harris}, Charles R. and {Millman}, K. Jarrod and {van der Walt}, St{\'e}fan J. and {Gommers}, Ralf and {Virtanen}, Pauli and {Cournapeau}, David and {Wieser}, Eric and {Taylor}, Julian and {Berg}, Sebastian and {Smith}, Nathaniel J. and {Kern}, Robert and {Picus}, Matti and {Hoyer}, Stephan and {van Kerkwijk}, Marten H. and {Brett}, Matthew and {Haldane}, Allan and {del R{\'\i}o}, Jaime Fern{\'a}ndez and {Wiebe}, Mark and {Peterson}, Pearu and {G{\'e}rard-Marchant}, Pierre and {Sheppard}, Kevin and {Reddy}, Tyler and {Weckesser}, Warren and {Abbasi}, Hameer and {Gohlke}, Christoph and {Oliphant}, Travis E.},
        title = "{Array programming with NumPy}",
      journal = {\nat},
         year = 2020,
        month = sep,
       volume = {585},
       number = {7825},
        pages = {357-362},
          doi = {10.1038/s41586-020-2649-2},
archivePrefix = {arXiv},
       eprint = {2006.10256},
 primaryClass = {cs.MS},
       adsurl = {https://ui.adsabs.harvard.edu/abs/2020Natur.585..357H}
}

@Article{Matplotlib,
  Author    = {Hunter, J. D.},
  Title     = {Matplotlib: A 2D graphics environment},
  Journal   = {Computing in Science \& Engineering},
  Volume    = {9},
  Number    = {3},
  Pages     = {90--95},
  publisher = {IEEE COMPUTER SOC},
  doi       = {10.1109/MCSE.2007.55},
  year      = 2007
}

@misc{pandas,
    author       = {The pandas development team},
    title        = {pandas-dev/pandas: Pandas},
    month        = feb,
    year         = 2020,
    publisher    = {Zenodo},
    version      = {latest},
    doi          = {10.5281/zenodo.3509134},
    url          = {https://doi.org/10.5281/zenodo.3509134},
    urldate = {12 Jan 2024}
}

@InProceedings{pandas2010,
  author    = { {W}es {M}c{K}inney },
  title     = { {D}ata {S}tructures for {S}tatistical {C}omputing in {P}ython },
  booktitle = { {P}roceedings of the 9th {P}ython in {S}cience {C}onference },
  pages     = { 56 - 61 },
  year      = { 2010 },
  editor    = { {S}t\'efan van der {W}alt and {J}arrod {M}illman },
  doi       = { 10.25080/Majora-92bf1922-00a }
}

@ARTICLE{astropy2013,
       author = {{Astropy Collaboration} and {Robitaille}, Thomas P. and {Tollerud}, Erik J. and {Greenfield}, Perry and {Droettboom}, Michael and {Bray}, Erik and {Aldcroft}, Tom and {Davis}, Matt and {Ginsburg}, Adam and {Price-Whelan}, Adrian M. and {Kerzendorf}, Wolfgang E. and {Conley}, Alexander and {Crighton}, Neil and {Barbary}, Kyle and {Muna}, Demitri and {Ferguson}, Henry and {Grollier}, Fr{\'e}d{\'e}ric and {Parikh}, Madhura M. and {Nair}, Prasanth H. and {Unther}, Hans M. and {Deil}, Christoph and {Woillez}, Julien and {Conseil}, Simon and {Kramer}, Roban and {Turner}, James E.~H. and {Singer}, Leo and {Fox}, Ryan and {Weaver}, Benjamin A. and {Zabalza}, Victor and {Edwards}, Zachary I. and {Azalee Bostroem}, K. and {Burke}, D.~J. and {Casey}, Andrew R. and {Crawford}, Steven M. and {Dencheva}, Nadia and {Ely}, Justin and {Jenness}, Tim and {Labrie}, Kathleen and {Lim}, Pey Lian and {Pierfederici}, Francesco and {Pontzen}, Andrew and {Ptak}, Andy and {Refsdal}, Brian and {Servillat}, Mathieu and {Streicher}, Ole},
        title = "{Astropy: A community Python package for astronomy}",
      journal = {\aap},
         year = 2013,
        month = oct,
       volume = {558},
          eid = {A33},
        pages = {A33},
          doi = {10.1051/0004-6361/201322068},
archivePrefix = {arXiv},
       eprint = {1307.6212},
 primaryClass = {astro-ph.IM},
       adsurl = {https://ui.adsabs.harvard.edu/abs/2013A&A...558A..33A}
}

@ARTICLE{astropy2018,
       author = {{Astropy Collaboration} and {Price-Whelan}, A.~M. and {Sip{\H{o}}cz}, B.~M. and {G{\"u}nther}, H.~M. and {Lim}, P.~L. and {Crawford}, S.~M. and {Conseil}, S. and {Shupe}, D.~L. and {Craig}, M.~W. and {Dencheva}, N. and {Ginsburg}, A. and {VanderPlas}, J.~T. and {Bradley}, L.~D. and {P{\'e}rez-Su{\'a}rez}, D. and {de Val-Borro}, M. and {Aldcroft}, T.~L. and {Cruz}, K.~L. and {Robitaille}, T.~P. and {Tollerud}, E.~J. and {Ardelean}, C. and {Babej}, T. and {Bach}, Y.~P. and {Bachetti}, M. and {Bakanov}, A.~V. and {Bamford}, S.~P. and {Barentsen}, G. and {Barmby}, P. and {Baumbach}, A. and {Berry}, K.~L. and {Biscani}, F. and {Boquien}, M. and {Bostroem}, K.~A. and {Bouma}, L.~G. and {Brammer}, G.~B. and {Bray}, E.~M. and {Breytenbach}, H. and {Buddelmeijer}, H. and {Burke}, D.~J. and {Calderone}, G. and {Cano Rodr{\'\i}guez}, J.~L. and {Cara}, M. and {Cardoso}, J.~V.~M. and {Cheedella}, S. and {Copin}, Y. and {Corrales}, L. and {Crichton}, D. and {D'Avella}, D. and {Deil}, C. and {Depagne}, {\'E}. and {Dietrich}, J.~P. and {Donath}, A. and {Droettboom}, M. and {Earl}, N. and {Erben}, T. and {Fabbro}, S. and {Ferreira}, L.~A. and {Finethy}, T. and {Fox}, R.~T. and {Garrison}, L.~H. and {Gibbons}, S.~L.~J. and {Goldstein}, D.~A. and {Gommers}, R. and {Greco}, J.~P. and {Greenfield}, P. and {Groener}, A.~M. and {Grollier}, F. and {Hagen}, A. and {Hirst}, P. and {Homeier}, D. and {Horton}, A.~J. and {Hosseinzadeh}, G. and {Hu}, L. and {Hunkeler}, J.~S. and {Ivezi{\'c}}, {\v{Z}}. and {Jain}, A. and {Jenness}, T. and {Kanarek}, G. and {Kendrew}, S. and {Kern}, N.~S. and {Kerzendorf}, W.~E. and {Khvalko}, A. and {King}, J. and {Kirkby}, D. and {Kulkarni}, A.~M. and {Kumar}, A. and {Lee}, A. and {Lenz}, D. and {Littlefair}, S.~P. and {Ma}, Z. and {Macleod}, D.~M. and {Mastropietro}, M. and {McCully}, C. and {Montagnac}, S. and {Morris}, B.~M. and {Mueller}, M. and {Mumford}, S.~J. and {Muna}, D. and {Murphy}, N.~A. and {Nelson}, S. and {Nguyen}, G.~H. and {Ninan}, J.~P. and {N{\"o}the}, M. and {Ogaz}, S. and {Oh}, S. and {Parejko}, J.~K. and {Parley}, N. and {Pascual}, S. and {Patil}, R. and {Patil}, A.~A. and {Plunkett}, A.~L. and {Prochaska}, J.~X. and {Rastogi}, T. and {Reddy Janga}, V. and {Sabater}, J. and {Sakurikar}, P. and {Seifert}, M. and {Sherbert}, L.~E. and {Sherwood-Taylor}, H. and {Shih}, A.~Y. and {Sick}, J. and {Silbiger}, M.~T. and {Singanamalla}, S. and {Singer}, L.~P. and {Sladen}, P.~H. and {Sooley}, K.~A. and {Sornarajah}, S. and {Streicher}, O. and {Teuben}, P. and {Thomas}, S.~W. and {Tremblay}, G.~R. and {Turner}, J.~E.~H. and {Terr{\'o}n}, V. and {van Kerkwijk}, M.~H. and {de la Vega}, A. and {Watkins}, L.~L. and {Weaver}, B.~A. and {Whitmore}, J.~B. and {Woillez}, J. and {Zabalza}, V. and {Astropy Contributors}},
        title = "{The Astropy Project: Building an Open-science Project and Status of the v2.0 Core Package}",
      journal = {\aj},
         year = 2018,
        month = sep,
       volume = {156},
       number = {3},
          eid = {123},
        pages = {123},
          doi = {10.3847/1538-3881/aabc4f},
archivePrefix = {arXiv},
       eprint = {1801.02634},
 primaryClass = {astro-ph.IM},
       adsurl = {https://ui.adsabs.harvard.edu/abs/2018AJ....156..123A}
}

@ARTICLE{astropy2022,
       author = {{Astropy Collaboration} and {Price-Whelan}, Adrian M. and {Lim}, Pey Lian and {Earl}, Nicholas and {Starkman}, Nathaniel and {Bradley}, Larry and {Shupe}, David L. and {Patil}, Aarya A. and {Corrales}, Lia and {Brasseur}, C.~E. and {N{\"o}the}, Maximilian and {Donath}, Axel and {Tollerud}, Erik and {Morris}, Brett M. and {Ginsburg}, Adam and {Vaher}, Eero and {Weaver}, Benjamin A. and {Tocknell}, James and {Jamieson}, William and {van Kerkwijk}, Marten H. and {Robitaille}, Thomas P. and {Merry}, Bruce and {Bachetti}, Matteo and {G{\"u}nther}, H. Moritz and {Aldcroft}, Thomas L. and {Alvarado-Montes}, Jaime A. and {Archibald}, Anne M. and {B{\'o}di}, Attila and {Bapat}, Shreyas and {Barentsen}, Geert and {Baz{\'a}n}, Juanjo and {Biswas}, Manish and {Boquien}, M{\'e}d{\'e}ric and {Burke}, D.~J. and {Cara}, Daria and {Cara}, Mihai and {Conroy}, Kyle E. and {Conseil}, Simon and {Craig}, Matthew W. and {Cross}, Robert M. and {Cruz}, Kelle L. and {D'Eugenio}, Francesco and {Dencheva}, Nadia and {Devillepoix}, Hadrien A.~R. and {Dietrich}, J{\"o}rg P. and {Eigenbrot}, Arthur Davis and {Erben}, Thomas and {Ferreira}, Leonardo and {Foreman-Mackey}, Daniel and {Fox}, Ryan and {Freij}, Nabil and {Garg}, Suyog and {Geda}, Robel and {Glattly}, Lauren and {Gondhalekar}, Yash and {Gordon}, Karl D. and {Grant}, David and {Greenfield}, Perry and {Groener}, Austen M. and {Guest}, Steve and {Gurovich}, Sebastian and {Handberg}, Rasmus and {Hart}, Akeem and {Hatfield-Dodds}, Zac and {Homeier}, Derek and {Hosseinzadeh}, Griffin and {Jenness}, Tim and {Jones}, Craig K. and {Joseph}, Prajwel and {Kalmbach}, J. Bryce and {Karamehmetoglu}, Emir and {Ka{\l}uszy{\'n}ski}, Miko{\l}aj and {Kelley}, Michael S.~P. and {Kern}, Nicholas and {Kerzendorf}, Wolfgang E. and {Koch}, Eric W. and {Kulumani}, Shankar and {Lee}, Antony and {Ly}, Chun and {Ma}, Zhiyuan and {MacBride}, Conor and {Maljaars}, Jakob M. and {Muna}, Demitri and {Murphy}, N.~A. and {Norman}, Henrik and {O'Steen}, Richard and {Oman}, Kyle A. and {Pacifici}, Camilla and {Pascual}, Sergio and {Pascual-Granado}, J. and {Patil}, Rohit R. and {Perren}, Gabriel I. and {Pickering}, Timothy E. and {Rastogi}, Tanuj and {Roulston}, Benjamin R. and {Ryan}, Daniel F. and {Rykoff}, Eli S. and {Sabater}, Jose and {Sakurikar}, Parikshit and {Salgado}, Jes{\'u}s and {Sanghi}, Aniket and {Saunders}, Nicholas and {Savchenko}, Volodymyr and {Schwardt}, Ludwig and {Seifert-Eckert}, Michael and {Shih}, Albert Y. and {Jain}, Anany Shrey and {Shukla}, Gyanendra and {Sick}, Jonathan and {Simpson}, Chris and {Singanamalla}, Sudheesh and {Singer}, Leo P. and {Singhal}, Jaladh and {Sinha}, Manodeep and {Sip{\H{o}}cz}, Brigitta M. and {Spitler}, Lee R. and {Stansby}, David and {Streicher}, Ole and {{\v{S}}umak}, Jani and {Swinbank}, John D. and {Taranu}, Dan S. and {Tewary}, Nikita and {Tremblay}, Grant R. and {de Val-Borro}, Miguel and {Van Kooten}, Samuel J. and {Vasovi{\'c}}, Zlatan and {Verma}, Shresth and {de Miranda Cardoso}, Jos{\'e} Vin{\'\i}cius and {Williams}, Peter K.~G. and {Wilson}, Tom J. and {Winkel}, Benjamin and {Wood-Vasey}, W.~M. and {Xue}, Rui and {Yoachim}, Peter and {Zhang}, Chen and {Zonca}, Andrea and {Astropy Project Contributors}},
        title = "{The Astropy Project: Sustaining and Growing a Community-oriented Open-source Project and the Latest Major Release (v5.0) of the Core Package}",
      journal = {\apj},
         year = 2022,
        month = aug,
       volume = {935},
       number = {2},
          eid = {167},
        pages = {167},
          doi = {10.3847/1538-4357/ac7c74},
archivePrefix = {arXiv},
       eprint = {2206.14220},
 primaryClass = {astro-ph.IM},
       adsurl = {https://ui.adsabs.harvard.edu/abs/2022ApJ...935..167A}
}

@article{seaborn,
    doi = {10.21105/joss.03021},
    url = {https://doi.org/10.21105/joss.03021},
    year = {2021},
    publisher = {The Open Journal},
    volume = {6},
    number = {60},
    pages = {3021},
    author = {Michael L. Waskom},
    title = {seaborn: statistical data visualization},
    journal = {Journal of Open Source Software}
 }

@ARTICLE{ray,
       author = {{Moritz}, Philipp and {Nishihara}, Robert and {Wang}, Stephanie and {Tumanov}, Alexey and {Liaw}, Richard and {Liang}, Eric and {Elibol}, Melih and {Yang}, Zongheng and {Paul}, William and {Jordan}, Michael I. and {Stoica}, Ion},
        title = "{Ray: A Distributed Framework for Emerging AI Applications}",
      journal = {arXiv e-prints},
         year = 2017,
        month = dec,
          eid = {arXiv:1712.05889},
        pages = {arXiv:1712.05889},
          doi = {10.48550/arXiv.1712.05889},
archivePrefix = {arXiv},
       eprint = {1712.05889},
 primaryClass = {cs.DC},
       adsurl = {https://ui.adsabs.harvard.edu/abs/2017arXiv171205889M}
}

@ARTICLE{astroquery,
       author = {{Ginsburg}, Adam and {Sip{\H{o}}cz}, Brigitta M. and {Brasseur}, C.~E. and {Cowperthwaite}, Philip S. and {Craig}, Matthew W. and {Deil}, Christoph and {Guillochon}, James and {Guzman}, Giannina and {Liedtke}, Simon and {Lian Lim}, Pey and {Lockhart}, Kelly E. and {Mommert}, Michael and {Morris}, Brett M. and {Norman}, Henrik and {Parikh}, Madhura and {Persson}, Magnus V. and {Robitaille}, Thomas P. and {Segovia}, Juan-Carlos and {Singer}, Leo P. and {Tollerud}, Erik J. and {de Val-Borro}, Miguel and {Valtchanov}, Ivan and {Woillez}, Julien and {Astroquery Collaboration} and {a subset of astropy Collaboration}},
        title = "{astroquery: An Astronomical Web-querying Package in Python}",
      journal = {\aj},
         year = 2019,
        month = mar,
       volume = {157},
       number = {3},
          eid = {98},
        pages = {98},
          doi = {10.3847/1538-3881/aafc33},
archivePrefix = {arXiv},
       eprint = {1901.04520},
 primaryClass = {astro-ph.IM},
       adsurl = {https://ui.adsabs.harvard.edu/abs/2019AJ....157...98G}
}

@ARTICLE{Euclid,
       author = {{Laureijs}, R. and {Amiaux}, J. and {Arduini}, S. and {Augu{\`e}res}, J. -L. and {Brinchmann}, J. and {Cole}, R. and {Cropper}, M. and {Dabin}, C. and {Duvet}, L. and {Ealet}, A. and {Garilli}, B. and {Gondoin}, P. and {Guzzo}, L. and {Hoar}, J. and {Hoekstra}, H. and {Holmes}, R. and {Kitching}, T. and {Maciaszek}, T. and {Mellier}, Y. and {Pasian}, F. and {Percival}, W. and {Rhodes}, J. and {Saavedra Criado}, G. and {Sauvage}, M. and {Scaramella}, R. and {Valenziano}, L. and {Warren}, S. and {Bender}, R. and {Castander}, F. and {Cimatti}, A. and {Le F{\`e}vre}, O. and {Kurki-Suonio}, H. and {Levi}, M. and {Lilje}, P. and {Meylan}, G. and {Nichol}, R. and {Pedersen}, K. and {Popa}, V. and {Rebolo Lopez}, R. and {Rix}, H. -W. and {Rottgering}, H. and {Zeilinger}, W. and {Grupp}, F. and {Hudelot}, P. and {Massey}, R. and {Meneghetti}, M. and {Miller}, L. and {Paltani}, S. and {Paulin-Henriksson}, S. and {Pires}, S. and {Saxton}, C. and {Schrabback}, T. and {Seidel}, G. and {Walsh}, J. and {Aghanim}, N. and {Amendola}, L. and {Bartlett}, J. and {Baccigalupi}, C. and {Beaulieu}, J. -P. and {Benabed}, K. and {Cuby}, J. -G. and {Elbaz}, D. and {Fosalba}, P. and {Gavazzi}, G. and {Helmi}, A. and {Hook}, I. and {Irwin}, M. and {Kneib}, J. -P. and {Kunz}, M. and {Mannucci}, F. and {Moscardini}, L. and {Tao}, C. and {Teyssier}, R. and {Weller}, J. and {Zamorani}, G. and {Zapatero Osorio}, M.~R. and {Boulade}, O. and {Foumond}, J.~J. and {Di Giorgio}, A. and {Guttridge}, P. and {James}, A. and {Kemp}, M. and {Martignac}, J. and {Spencer}, A. and {Walton}, D. and {Bl{\"u}mchen}, T. and {Bonoli}, C. and {Bortoletto}, F. and {Cerna}, C. and {Corcione}, L. and {Fabron}, C. and {Jahnke}, K. and {Ligori}, S. and {Madrid}, F. and {Martin}, L. and {Morgante}, G. and {Pamplona}, T. and {Prieto}, E. and {Riva}, M. and {Toledo}, R. and {Trifoglio}, M. and {Zerbi}, F. and {Abdalla}, F. and {Douspis}, M. and {Grenet}, C. and {Borgani}, S. and {Bouwens}, R. and {Courbin}, F. and {Delouis}, J. -M. and {Dubath}, P. and {Fontana}, A. and {Frailis}, M. and {Grazian}, A. and {Koppenh{\"o}fer}, J. and {Mansutti}, O. and {Melchior}, M. and {Mignoli}, M. and {Mohr}, J. and {Neissner}, C. and {Noddle}, K. and {Poncet}, M. and {Scodeggio}, M. and {Serrano}, S. and {Shane}, N. and {Starck}, J. -L. and {Surace}, C. and {Taylor}, A. and {Verdoes-Kleijn}, G. and {Vuerli}, C. and {Williams}, O.~R. and {Zacchei}, A. and {Altieri}, B. and {Escudero Sanz}, I. and {Kohley}, R. and {Oosterbroek}, T. and {Astier}, P. and {Bacon}, D. and {Bardelli}, S. and {Baugh}, C. and {Bellagamba}, F. and {Benoist}, C. and {Bianchi}, D. and {Biviano}, A. and {Branchini}, E. and {Carbone}, C. and {Cardone}, V. and {Clements}, D. and {Colombi}, S. and {Conselice}, C. and {Cresci}, G. and {Deacon}, N. and {Dunlop}, J. and {Fedeli}, C. and {Fontanot}, F. and {Franzetti}, P. and {Giocoli}, C. and {Garcia-Bellido}, J. and {Gow}, J. and {Heavens}, A. and {Hewett}, P. and {Heymans}, C. and {Holland}, A. and {Huang}, Z. and {Ilbert}, O. and {Joachimi}, B. and {Jennins}, E. and {Kerins}, E. and {Kiessling}, A. and {Kirk}, D. and {Kotak}, R. and {Krause}, O. and {Lahav}, O. and {van Leeuwen}, F. and {Lesgourgues}, J. and {Lombardi}, M. and {Magliocchetti}, M. and {Maguire}, K. and {Majerotto}, E. and {Maoli}, R. and {Marulli}, F. and {Maurogordato}, S. and {McCracken}, H. and {McLure}, R. and {Melchiorri}, A. and {Merson}, A. and {Moresco}, M. and {Nonino}, M. and {Norberg}, P. and {Peacock}, J. and {Pello}, R. and {Penny}, M. and {Pettorino}, V. and {Di Porto}, C. and {Pozzetti}, L. and {Quercellini}, C. and {Radovich}, M. and {Rassat}, A. and {Roche}, N. and {Ronayette}, S. and {Rossetti}, E. and {Sartoris}, B. and {Schneider}, P. and {Semboloni}, E. and {Serjeant}, S. and {Simpson}, F. and {Skordis}, C. and {Smadja}, G. and {Smartt}, S. and {Spano}, P. and {Spiro}, S. and {Sullivan}, M. and {Tilquin}, A. and {Trotta}, R. and {Verde}, L. and {Wang}, Y. and {Williger}, G. and {Zhao}, G. and {Zoubian}, J. and {Zucca}, E.},
        title = "{Euclid Definition Study Report}",
      journal = {arXiv e-prints},
         year = 2011,
        month = oct,
          eid = {arXiv:1110.3193},
        pages = {arXiv:1110.3193},
          doi = {10.48550/arXiv.1110.3193},
archivePrefix = {arXiv},
       eprint = {1110.3193},
 primaryClass = {astro-ph.CO},
       adsurl = {https://ui.adsabs.harvard.edu/abs/2011arXiv1110.3193L}
}

@ARTICLE{Mannucci2022,
       author = {{Mannucci}, F. and {Pancino}, E. and {Belfiore}, F. and {Cicone}, C. and {Ciurlo}, A. and {Cresci}, G. and {Lusso}, E. and {Marasco}, A. and {Marconi}, A. and {Nardini}, E. and {Pinna}, E. and {Severgnini}, P. and {Saracco}, P. and {Tozzi}, G. and {Yeh}, S.},
        title = "{Unveiling the population of dual and lensed active galactic nuclei at sub-arcsec separations}",
      journal = {Nature Astronomy},
         year = 2022,
        month = aug,
       volume = {6},
        pages = {1185-1192},
          doi = {10.1038/s41550-022-01761-5},
archivePrefix = {arXiv},
       eprint = {2203.11234},
 primaryClass = {astro-ph.GA},
       adsurl = {https://ui.adsabs.harvard.edu/abs/2022NatAs...6.1185M}
}

@misc{VATLoss,
  author = {Lyakaap},
  title = {VAT-pytorch},
  year = 2018,
  url = {https://github.com/lyakaap/VAT-pytorch},
  urldate = {2024-12-15}
}

@misc{zenodo-code,
  author       = {David Sweeney},
  title        = {Code supplement to Semi-Supervised Learning for Lensed Quasar Detection},
  month        = dec,
  year         = 2025,
  publisher    = {Zenodo},
  version      = {1.0.0},
  doi          = {10.5281/zenodo.14490178},
  url          = {https://doi.org/10.5281/zenodo.14490178},
  urldate = {5 Dec 2025}
}

@ARTICLE{Oke1995,
       author = {{Oke}, J.~B. and {Cohen}, J.~G. and {Carr}, M. and {Cromer}, J. and {Dingizian}, A. and {Harris}, F.~H. and {Labrecque}, S. and {Lucinio}, R. and {Schaal}, W. and {Epps}, H. and {Miller}, J.},
        title = "{The Keck Low-Resolution Imaging Spectrometer}",
      journal = {\pasp},
         year = 1995,
        month = apr,
       volume = {107},
        pages = {375},
          doi = {10.1086/133562},
       adsurl = {https://ui.adsabs.harvard.edu/abs/1995PASP..107..375O}
}

@ARTICLE{Krone-Martins2018,
       author = {{Krone-Martins}, A. and {Delchambre}, L. and {Wertz}, O. and {Ducourant}, C. and {Mignard}, F. and {Teixeira}, R. and {Kl{\"u}ter}, J. and {Le Campion}, J. -F. and {Galluccio}, L. and {Surdej}, J. and {Bastian}, U. and {Wambsganss}, J. and {Graham}, M.~J. and {Djorgovski}, S.~G. and {Slezak}, E.},
        title = "{Gaia GraL: Gaia DR2 gravitational lens systems. I. New quadruply imaged quasar candidates around known quasars}",
      journal = {\aap},
         year = 2018,
        month = aug,
       volume = {616},
          eid = {L11},
        pages = {L11},
          doi = {10.1051/0004-6361/201833337},
archivePrefix = {arXiv},
       eprint = {1804.11051},
 primaryClass = {astro-ph.GA},
       adsurl = {https://ui.adsabs.harvard.edu/abs/2018A&A...616L..11K}
}

% Alternatively you could enter them by hand, like this:
% This method is tedious and prone to error if you have lots of references
%\begin{thebibliography}{99}
%\bibitem[\protect\citeauthoryear{Author}{2012}]{Author2012}
%Author A.~N., 2013, Journal of Improbable Astronomy, 1, 1
%\bibitem[\protect\citeauthoryear{Others}{2013}]{Others2013}
%Others S., 2012, Journal of Interesting Stuff, 17, 198
%\end{thebibliography}

%%%%%%%%%%%%%%%%%%%%%%%%%%%%%%%%%%%%%%%%%%%%%%%%%%

%%%%%%%%%%%%%%%%% APPENDICES %%%%%%%%%%%%%%%%%%%%%

\appendix

\section{Pan-STARRS1 JPEG Compression}
\label{app:JPEG}
The Pan-STARRS1 cutout images are generated via a two-step process, here I paraphrase the relevant section of an email from Rick White (priv. comm., 2025) from the Pan-STARRS1 data processing team describing this process:

\begin{enumerate}
    \item ``The image contrast is limited using a linear stretch between the 0.5\% and 99.5\% pixel values within the cutout. This process clips the occasional bad pixel or bright star. 
    \item Then an arc-hyperbolic sine (asinh) transform is applied to the scaled pixels values. The asinh function behaves similarly to the log function for large values ($\gg 1$ or $\ll -1$) but is linear for values near zero. It is also (unlike log) defined for negative pixels. This behaviour enhances contrast near the sky and further suppresses the brightness of bright stars.
\end{enumerate}

The underlying C code for the actual equations for the contrast computation can be found here: \url{https://github.com/spacetelescope/fitscut}''

The clipping of bad pixels and asinh transformation of the pixel values likely result in the increased performance observed in this work.

\section{F1, Recall and Precision Scores}
\label{app:scores}
A precision score is defined as follows:
\begin{align}
    \text{Precision} &= \frac{\text{True positives}}{\text{True positives + False positives}}
\end{align}
A recall score is defined as follows:
\begin{align}
    \text{Recall} &= \frac{\text{True positives}}{\text{True positives + False negatives}}
\end{align}
An F1 score is the harmonic mean of precision and recall:
\begin{align}
    \text{F}_1 &= 2\cdot\frac{\text{Precision}\cdot\text{Recall}}{\text{Precision + Recall}}
\end{align}

\section{Tuning the classifiers}
\label{app:classifiers}

The random forest models were trained with an entropy criterion and the hyperparameters ranged over 50--500 estimators, 1--20 minimum samples per leaf, 1--all features considered in each subtree and the weighting of the lensed and unlensed classes being ``balanced'' or ``balanced\_subsample''.

The logistic regression models were trained with an elastic net penalty, a ``balanced'' class weight, a maxiumum number of iterations of 1000 and a saga solver. The hyperparameter search spanned l1\_ratios of [0, 0.1, 0.3, 0.5, 0.7, 0.9, 0.95, 0.99, 1]. These values allowed exploration of models ranging from pure L2 regularization (l1\_ratio = 0) to pure L1 regularization (l1\_ratio = 1), as well as intermediate combinations of the two regularization techniques.

The gradient boosting models were trained with a hyperparameter search over the number of estimators was varied between 50 and 500, the learning rate between 0.1 and 0.5. The maximum depth of the trees ranged from 10 to 50, and the subsample fraction spanned values from 0.5 to 1.0. Additionally, the class imbalance was addressed by tuning the `scale\_pos\_weight` parameter over from 56 to 316, enabling the model to effectively handle varying levels of class weighting.

The linear SVM was also trained using a ``balanced'' class weight, a squared hinge loss and maximum of 10$\,$000 iterations. The hyperparameter search tried an L1 and L2 (but not both) with a regularisation parameter ``C'' ranging from 0.1 to 10$\,$000.

The kernel-SVM model was trained with a ``balanced'' class weight and tried both a polynomial and a radial basis function kernels. The hyperparameter search for the regularisation parameter ranged from $10^{-4}$--$10^4$, the degree of the polynomial kernel from 2--5 and the kernel coefficient was either ``scale'' or ``auto''.

The isolation forests were trained with an ``auto'' contamination. The hyperparameter search for the number of estimators ranged from 300--400, the number of samples from 2/5--4/5 of the available data, the number of features from 2/5--1/2 of the available and bootstrapping being used or not used. 

The Gaussian processes simply had 50 restarts for the optimiser.

The artificial neural network wasn't tuned using {\tt RandomizedSearchCV} as the custom implementation didn't easily fit into this framework.  Instead, we used {\tt StratifiedKFold} and a grid search over the following sets of hidden layers: (64, 64), (64, 32), (32, 32), (64, 64, 64), (64, 32, 16), (32, 32, 32), (64, 64, 64, 64), (64, 32, 16, 8) and (32, 32, 32, 32), combined with either dropout (tested values of 0.05, 0.1, 0.125, 0.15, 0.175 and 0.2) or weight decay ($10^{-4}$, $3\cdot 10^{-3}$, $10^{-3}$, $3\cdot10^{-2}$, $10^{-2}$) with PyTorch's {\tt BCEWithLogitsLoss} as the loss function. Because of the variation model performance due to the randomised initial weights and biases of the neural network, the best combination of hidden layers and regularisation varied on consecutive runs. To find the best set of hyperparameters {\tt StratifiedKFold} run 7 times with seeds from 0--6 and the best model was selected based on the following criteria:

\begin{enumerate}
    \item The set of hidden layers which appeared the most times as the best set of hidden layers.
    \item If the regularisation differs amongst solutions, the value which appears most frequently is chosen.
    \item Ties are broken by the lowest average binary cross entropy error.
\end{enumerate}

The code used to tune the models is available on Zenodo \citep{zenodo-code}.

%%%%%%%%%%%%%%%%%%%%%%%%%%%%%%%%%%%%%%%%%%%%%%%%%%

% Don't change these lines
\bsp	% typesetting comment
\label{lastpage}
\end{document}